\documentclass[a4paper,fleqn,usenatbib]{mn2e}

\usepackage{amssymb}
\usepackage{amsmath}
\usepackage{txfonts}
\usepackage[T1]{fontenc}
\usepackage{times}
\usepackage{BibDef}
\usepackage{hyperref}
\usepackage{float}
\usepackage{color}
\usepackage{subfigure}
\usepackage{graphicx}
\usepackage{soul}

\let\oldAA\AA
\renewcommand{\AA}{\text{\normalfont\oldAA}}

\newcommand{\ESD}                         {$\Delta\Sigma\,\,$}

\newcommand{\Mpc}                      {\,{\rm Mpc}}
\newcommand{\Msun}                    {\,{\rm M}_\odot}

\newcommand{\hkpc}                     {\,{\rm h}^{-1}\,{\rm kpc}}
\newcommand{\hMpc}                    {\,{\rm h}^{-1}\,{\rm Mpc}}

\newcommand{\newl} 	{\mathrm{log_{10}}}

\title[Galaxy-Galaxy Lensing in EAGLE]{Galaxy-Galaxy Lensing in EAGLE: comparison with data from 180 square degrees of the KiDS and GAMA surveys}
\author[Marco Velliscig et al.]
{Marco Velliscig,$^{1}$\thanks{E-mail: velliscig@strw.leidenuniv.nl} 
Marcello Cacciato,$^{1}$ 
Henk Hoekstra,$^{1}$ 
Joop Schaye,$^{1}$ 
\newauthor
Catherine Heymans,$^{2}$
Hendrik Hildebrandt,$^{3}$
Jon Loveday,$^{4}$
Peder Norberg,$^{5}$
\newauthor
Crist\'obal Sif\'on,$^{1,6}$
Peter Schneider,$^{3}$
Edo van Uitert,$^{7}$ 
Massimo Viola,$^{1}$
Sarah Brough,$^{8}$
\newauthor
Thomas Erben,$^{3}$
Benne W. Holwerda,$^{1}$
Andrew M. Hopkins,$^{8}$
Konrad Kuijken$^{1}$ 
\\
\\
$^{1}$Leiden Observatory, Leiden University, P.O. Box 9513, 2300 RA Leiden, The Netherlands\\
$^{2}$Scottish Universities Physics Alliance, Institute for Astronomy, University of Edinburgh, Royal Observatory, Blackford Hill, Edinburgh EH9 3HJ, UK\\
$^{3}$Argelander-Institut f{\"u}r Astronomie, Auf dem H{\"u}gel 71, D-53121 Bonn, Germany\\
$^{4}$Astronomy Centre, Department of Physics and Astronomy, University of Sussex, Falmer, Brighton BN1 9QH, UK\\
$^{5}$ICC and CEA, Department of Physics, Durham University, South Road, Durham DH1 3LE, UK\\
$^{6}$Department of Astrophysical Sciences, Peyton Hall, Princeton University, Princeton, NJ 08544, USA\\
$^{7}$University College London, Gower Street, London WC1E 6BT, UK\\
$^{8}$Australian Astronomical Observatory, PO Box 915, North Ryde, NSW 1670, Australia
}

\date{Accepted XXX. Received YYY; in original form ZZZ}

\pubyear{2016}

\begin{document}
\label{firstpage}
\pagerange{\pageref{firstpage}--\pageref{lastpage}}
\maketitle

\begin{abstract}
We present predictions for the galaxy-galaxy lensing profile from the EAGLE hydrodynamical cosmological simulation at redshift $z$=0.18, 
in the spatial range $0.02<R/(h^{-1}\rm{Mpc})<2$, and for five logarithmically equi-spaced stellar mass bins in the range $10.3<\newl(M_{\rm star}/\Msun)< 11.8$. 
We compare these excess surface  density profiles to the observed signal from background galaxies imaged by the Kilo Degree Survey around spectroscopically confirmed foreground galaxies from the GAMA survey. Exploiting the GAMA galaxy group catalogue, the profiles of central and satellite galaxies are computed separately for groups with at least five members 
to minimise contamination. EAGLE predictions are in broad agreement with the observed profiles for both central and satellite galaxies, although the signal is underestimated at $R \approx 0.5-2 h^{-1}\rm{Mpc}$ for the highest stellar mass bins. When central and satellite galaxies are considered simultaneously, agreement is found only when the selection function of lens galaxies is taken into account in detail. Specifically, in the case of GAMA galaxies, it is crucial to account for the variation of the fraction of satellite galaxies in bins of stellar mass induced by the flux-limited nature of the survey. We report the inferred stellar-to-halo mass relation and we find good agreement with recent published results. We note how the precision of the galaxy-galaxy lensing profiles in the simulation holds the potential to constrain fine-grained aspects of the galaxy-dark matter connection. 
\end{abstract}


\section{Introduction}
\label{Sec:introduction}

The connection between observable galaxy properties and the underlying (mostly dark) matter density field is the result of galaxy formation and evolution in a cosmological context; 
as such, it is extensively studied from various complementary perspectives. Numerous methods are available to probe the mass of dark matter haloes within the galaxy formation 
framework, such as galaxy clustering \citep[see e.g.][]{Jing98, Peacock00,Zehavi02, vandenBosch03,Anderson14}, abundance matching \citep[see e.g.][]{Vale04, Moster13, Behroozi13} 
and stacked satellite kinematics \citep[see e.g.][]{Zaritsky94, Prada03, Conroy05, More11c}. These methods require, in various ways, prior knowledge of galaxy formation theory. 
They are, therefore, limited in their capacity to produce a stellar mass versus halo mass relation that can serve as a test for the galaxy formation framework itself.

For single galaxies, direct methods for estimating the halo mass are available \citep[see for a recent review][]{Courteau14}. 
The rotation curves of spiral galaxies or the velocity dispersions of ellipticals can give estimates of the amount of matter associated with a galaxy, 
albeit at relatively small scales. Furthermore, a galaxy can deflect the light of a background galaxy along the line of sight, possibly into multiple images,
providing a measurement of the total projected mass within the Einstein radii of galaxies \citep[][and references therein]{Kochanek91, Bolton08, Collett15}.  
The mass of a single group or cluster of galaxies can be estimated via the dynamics of its satellite galaxies \citep[see e.g.][]{Prada03, Conroy05}, 
using weak or strong lensing \citep[see e.g.][]{Hoekstra15, Fort94, Massey10} or X-ray emission \citep[][and references therein]{Ettori13}.

For a population of galaxies, galaxy-galaxy weak lensing \citep[see e.g.][]{Brainerd96, Wilson01, Hoekstra04, Mandelbaum06c, van_Uitert11, Velander14, Viola15, vanUitert15,Leauthaud15,Mandelbaum16}  
offers the possibility to measure the average halo mass directly and therefore represents a viable alternative to constrain the galaxy-dark matter connection 
and ultimately test galaxy formation models. Galaxy-galaxy lensing measures the distortion and magnification of the light of faint background galaxies (sources) 
caused by the deflection of light rays by intervening matter along the line of sight (lenses). The effect is independent of the dynamical state of the lens, and the projected mass 
of the lens is measured without any assumption about the physical state of the matter. The gravitational lensing signal due to a single galaxy is too weak to be detected 
(it is typically 10 to 100 times smaller than the intrinsic ellipticity of galaxies) given the typical number density of background sources in wide-field surveys. 
Therefore the galaxy-galaxy lensing signal must be  averaged over many lenses. 

From a more theoretical perspective, the link between haloes and galaxies can be studied with an {\it ab-initio} approach using Semi Analytical models 
and hydrodynamical cosmological simulations. Simulations aim to directly model the physical processes that are thought to be important for the formation of galaxies, 
as well as the energetic feedback from supernovae and AGN that is thought to regulate their growth \citep[see][for a recent review]{Somerville14}.
However, many of these processes are happening on scales that are unresolved by simulations and as such they must be treated as `subgrid' physics. 
To gain confidence in these physical recipes, it thus becomes crucial to compare predictions of these models to various observations. 
Arguably, a key test for such studies is to reproduce the observed abundances of galaxies as a  function of their stellar mass (galaxy stellar mass function; hereafter GSMF), 
as this is interpreted as the achievement of a successful mapping between the stellar mass and the halo mass. Intriguingly, reproducing a basic quantity such as the GSMF 
has proven to be extremely challenging for models of galaxy formation. To overcome this limitation, one might reverse the logic and calibrate the unresolved physical processes 
to reproduce the (present-day) GSMF. This approach, exploited at length in Semi Analytical models, has recently been adopted in hydro-simulations as well 
\citep[see e.g. the EAGLE and the BAHAMAS project, ][]{Schaye14,Crain15,McCarthy16}

In this paper, we compute the predicted weak galaxy-galaxy lensing (GGL) profiles of galaxies from the EAGLE hydrodynamical simulation sampled according to their stellar mass. 
These predictions are compared with the observed signal measured using background galaxies imaged by the Kilo Degree Survey \citep[KiDS; ][]{deJong15}
around spectroscopically confirmed foreground galaxies from the Galaxy And Mass Assembly (GAMA) survey \citep{Driver11}. We refer to this combined data set as KiDSxGAMA.
This comparison represents an  independent test of the validity of the physical processes implemented in the EAGLE simulation, as they were calibrated to reproduce 
the galaxy stellar mass function as well as the observed distribution of galaxy sizes but not the lensing profiles. As explained in the main body of the paper and in Appendix~A, 
a comparison of the GGL profiles offers the possibility to test fine-grained aspects of the galaxy-dark matter connection.

This paper is organized as follows. In Section~\ref{Sec:kids_gama} we briefly introduce the data sets and describe the methodology to obtain the galaxy-galaxy lensing measurements.
In Section \ref{Sec:Eagle} we describe the EAGLE simulation employed in this study, the algorithm used to produce the group catalogue from simulations (\S\ref{Sec:FOF_sim_ESD}) and the steps taken to measure the galaxy-galaxy lensing signal in the simulations (\S\ref{Sec:ESD_sim}). 
In Section \ref{Sec:Result_ESD} we report the results for the galaxy-galaxy lensing signal from simulations and the comparison with KiDSxGAMA data for central (\S\ref{Sec:ESD_sim_cen}) 
and satellites galaxies (\S\ref{Sec:ESD_sim_sat}). In \S\ref{Sec:ESD_sim_tot} we compare the GGL profile for the whole galaxy population against the KiDSxGAMA observations.
We discuss limitations and possible future improvements of this study in Section~\ref{Sec:discussion}, summarize our findings and conclude in Section~\ref{Sec:Conclusions_ESD}. 
We fit the galaxy-galaxy lensing profiles from the EAGLE simulation with simple analytical models in Appendix~A.

Throughout the paper we assume a $\Lambda$CDM cosmological model defined by the following set of parameters \{$\Omega_{ \rm m}$,\, $\Omega_{ \rm b}$,\,$\sigma_{ \rm 8}$,\, 
$n_{ \rm s}$,\, $h\equiv H_0/100$\} = \{0.307, 0.04825, 0.8288, 0.9611, 0.6777\} (motivated by the initial results from the Planck mission; \citealt{Planck13}), as this  was the cosmology assumed for the EAGLE run. We decided to maintain the explicit dependence on $h$ when plotting the galaxy-galaxy lensing profiles to ease the comparison with other published results.

\section{Data}
\label{Sec:kids_gama}

The  observational data presented in this paper are obtained from two  surveys: KiDS and GAMA. 
KiDS is an ESO optical imaging survey \citep{deJong13} with the OmegaCAM wide-field imager on the VLT Survey Telescope. 
When completed, it will cover a total area of 1500 square degrees in four bands ({\it u}, {\it g}, {\it r}, {\it i}).
KiDS was designed to have both good galaxy shape measurements and photometric redshift estimates of (background) galaxies. 
Here we use the latest KiDS-ESO data release which is described in \citet{Hildebrandt16}. Details of the survey can be found in \citet{deJong15}.

\begin{table*}
\centering
\begin{minipage}{168mm}
\begin{center}
\begin{tabular}{cccccccrcrcc}
\hline
$M_{\rm star}$ & $  M_{200}^{\rm crit}|_{\rm cen}$ & $\  M_{200}^{\rm crit}|_{\rm sat}$ & $\ M_{\rm sub}^{\rm cen}$ & $  M_{\rm sub}^{\rm sat}$ & $ M_{\rm sub}^{\rm sat}/M_{200}^{\rm crit}|_{\rm sat}$ & $ d_{\rm sat}$ & $  r_{\rm half}^{\rm dm}|_{\rm cen}$ & $  r_{\rm half}^{\rm dm}|_{\rm sat}$ & $N_{\rm gal}$ & $  M_{\rm star}^{\rm limit}$ & $  f_{\rm sat}$   \\
 * & * & * & * & * & & ** & ** & ** & & * &  \\ 
 (1)   &(2)            & (3)     & (4)    & (5)    & (6)    & (7)    &(8)  &(9)  &(10)&(11)&(12)\\
\hline
$[10.3-10.6]$ & $ 12.46 $ & $ 13.95 $ & $ 12.47 $ & $ 11.57 $ & $  0.03 $ & $ \hfill 881 $ & $ \hfill 144 $ & $ \hfill 28 $ & $ \hfill 354 $ & $ \hfill 9.46 $ & $ \hfill 0.98 $\\
$[10.6-10.9]$ & $ 12.92 $ & $ 14.09 $ & $ 12.92 $ & $ 11.95 $ & $  0.03 $ & $ \hfill 1081 $ & $ \hfill 239 $ & $ \hfill 44 $ & $ \hfill 150$ & $ \hfill 9.91 $ & $ \hfill 0.95 $\\
$[10.9-11.2]$ & $ 13.13 $ & $ 14.14 $ & $ 13.15 $ & $ 12.46 $ & $  0.11 $ & $ \hfill 1347 $ & $ \hfill 261 $ & $ \hfill 75 $ & $ \hfill 68 $ & $ \hfill 9.96 $ & $ \hfill 0.81 $\\
$[11.2-11.5]$ & $ 13.39 $ & $ 14.19 $ & $ 13.39 $ & $ 12.85 $ & $  0.13 $ & $ \hfill 1718 $ & $ \hfill 318 $ & $ \hfill 108 $ & $ \hfill 22 $ & $ \hfill 10.33 $ & $ \hfill 0.50 $\\
$[11.5-11.8]$ & $ 13.69 $ & $ 14.24 $ & $ 13.69 $ & $ 13.61 $ & $  0.30 $ & $ \hfill 2802 $ & $ \hfill 340 $ & $ \hfill 264 $ & $ \hfill 29 $ & $ -- $ & $ \hfill 0.21 $\\
\hline                                                   
\end{tabular}
\caption{ Various quantities of interest extracted from the EAGLE simulation at $z=0.18$. From left to right of the columns list: (1) stellar mass range; 
(2) average halo mass, $M_{200}^{\rm crit}$, of haloes hosting central galaxies in each stellar mass bin; 
(3) same as (2) but for haloes hosting satellites in each stellar mass bin;
(4) mean value of the subhalo mass for central galaxies, considering all the particles bound to the subhalo$^{***}$; 
(5) same as (4) but for satellite galaxies; 
(6) average ratio between the mass of the satellite subhalo, $M_{\rm sub}$, and the mass of its host halo $M_{200}^{\rm crit}$; 
(7) average 3D distance between the satellite galaxy and the centre of its host halo; 
(8) mean radius of central galaxies within which half of the mass in dark matter is enclosed; 
(9) same as (8) but for satellite galaxies; 
(10) total number of galaxies in the stellar mass bin; 
(11) Minimum stellar mass for which a galaxy is considered for the computation of the richness of its group in the EAGLE simulation. 
This value of $M_{\rm star}^{\rm limit}$ reproduces the satellite fraction in GAMA. Note that the value for the stellar mass bin $[11.5-11.8]$ is ill-defined (see discussion in \S 3.3). 
(12) average satellite fraction in EAGLE expressed as the total number of satellites divided by the total number of galaxies in the mass bin. 
This value is equal to the satellite fraction in GAMA by construction.
} 
\label{tbl_ch5:stat_ESD} 
\end{center}
\vspace{-0.2in}
\footnotetext{* $\newl (M/ [\Msun])$}
\footnotetext{** $R/[$kpc$]$}
\footnotetext{*** Note taht column (2) and (4) have very similar values. This indicates that, in this sample, adopting a spherical overdensity threshold or a FoF algorithm to define the halo yields to comparable halo masses.}
\end{minipage}
\end{table*}

KiDS overlaps with the GAMA spectroscopic survey \citep{Driver11} carried out using the AAOmega multi-object spectrograph on the Anglo-Australian Telescope (AAT). 
GAMA equatorial fields are 98\% complete down to a {\it r}-band magnitude of 19.8, and cover approximately $180$ sq. degrees of sky that fully overlap with the KiDS footprint. 
The redshift distribution of GAMA galaxies (median redshift $z\approx0.25$) is ideal for measurements of the galaxy-galaxy lensing signal 
using KiDS galaxies as background sources (median redshift $z\approx0.7$).

GAMA spectroscopy allows reliable identification of galaxy groups \citep[][]{Robotham11}, which in turn permits a separation between central and satellite galaxies. This distinction will be used extensively throughout the paper. 

\subsection{Lensing analysis}
\label{Sec:ESD_data}
A detailed description of how the galaxy-galaxy lensing signal around GAMA galaxies using KiDSxGAMA data is computed can be found in \citet{Viola15} and Dvornik et al. (in prep.). 
Here, we only summarize the important aspects that enter into the measurement.

Shape measurements are based on the {\it r}-band exposures which yield the highest image quality in KiDS.
Images are processed with the THELI pipeline \citep[optimized for lensing applications,][]{Erben05,Erben09,Erben13}, 
and galaxy ellipticities are computed using the \emph{lens}fit code \citep{Miller07, Kitching08, Miller13}. 
Shape measurements are calibrated against extensive image simulations \citep{Fenech16}.
Biases from non-perfect PSF modelling, are quantified and found subdominant as detailed in \citet{Kuijken15}.

For every lens-source pair, the measured ellipticity ($e_1$, $e_2$) of the source, as estimated by \emph{lens}fit, 
is projected along the separation of the lens in a tangential ($e_+$) and cross ($e_{\times}$) component as:
\begin{equation}
\begin{pmatrix} e_+ \\ e_{\times} \end{pmatrix}=\begin{pmatrix}-\cos(2\phi) & -\sin(2\phi) \\ \sin(2\phi) & -\cos(2\phi) \end{pmatrix}\begin{pmatrix} e_1 \\ e_2 \end{pmatrix},
\end{equation}
where $\phi$ is the angle between the x-axis and the lens-source separation vector.
Every source lens pair is then weighted by the term:
\begin{equation}
\tilde{w}_{\rm ls}=w_{\rm s}\, \langle \Sigma_{\rm crit}^{-1} \rangle_{\rm ls}^2
\end{equation}
which is the product of the \emph{lens}fit weight $w_{\rm s}$, computed according to the estimated reliability of the measured source ellipticity \citep{Miller07}, 
and a term $\langle \Sigma_{\rm crit}^{-1} \rangle_{\rm ls}^2$ defined via
\begin{equation}
\langle {\Sigma}_{\rm crit}^{-1} \rangle_{\rm ls} = \frac{4\pi G}{c^2}\, D_{\rm l}(z_{\rm l}) \,  \int\limits_{z_{\rm l}+\Delta z}^{\infty}\frac{D_{\rm ls}(z_{\rm l}, z_{\rm s})}{D_s(z_{\rm s})}
n(z_{\rm s}) {\rm d} z_{\rm s},
\end{equation}
where $D_{\rm l}$ is the angular diameter distance of the lens calculated using the spectroscopic redshift $z_{\rm l}$, $D_s$ is the angular diameter distance of the source,
and we have used $\Delta z = 0.2$ to minimize contamination by lenses (see Dvornik et al. in prep.). Here, $n(z_{\rm s})$ is the redshift distribution of the background galaxy population, 
and $D_{\rm ls}$ is the distance between the lens and the source. We emphasize here that $n(z_{\rm s})$ is the global redshift distribution of the KiDS galaxies 
estimated using the direct calibraton method described in \citet{Hildebrandt16}.

The galaxy-galaxy lensing signal, also known as the excess surface density, ESD, is computed in bins of projected distance $R$:
\begin{equation}
\label{Eq:ESD_data}
\Delta \Sigma(R) =\gamma_{\rm t}(R)\, \langle {\Sigma}_{\rm crit} \rangle_{\rm ls} = \left( \frac{\sum_{\rm ls}\tilde{w}_{\rm ls} e_{+} \langle \Sigma_{\rm crit} \rangle_{\rm ls}}{\sum_{\rm ls}\tilde{w}_{\rm ls}}\right) 
\frac{1}{1+K(R)}
\end{equation}
where $\langle {\Sigma}_{\rm crit} \rangle_{\rm ls} \equiv 1/\langle \Sigma_{\rm crit}^{-1}\rangle_{\rm ls}$. 
Here, the sum is over all lens-source pairs in the radial bin, and
\begin{equation}
K(R) = \frac{\sum_{\rm ls}\beta_{\rm ls} m_{\rm s}}{\sum_{\rm ls}\beta_{\rm ls}}
\end{equation}
is the correction to the ESD profile that takes into account the multiplicative bias $m_{\rm s}$, with $\beta_{\rm ls} = D_{\rm ls}/D_{\rm s}$. Typically, the value of $m_{\rm s}$ is around $-0.012$ which results in a $1 / (1+K(R))$ correction of $\sim 1.01$ (\citealt{Fenech16}, Dvornik et al. in prep.).

The error on the ESD measurement is estimated by:
\begin{equation}
\label{Eq:error_data}
\sigma^2_{\Delta\Sigma} = \sigma^2_{e_{+}}  \left( \frac{\sum_{\rm ls}\tilde{w}_{\rm ls}^2 \langle \Sigma_{\rm crit}^{-1}\rangle^2}{(\sum_{\rm ls}\tilde{w}_{\rm ls})^2}\right) ,
\end{equation}
where  $\sigma^2_{e_{+}}$ is the variance of all source ellipticities combined. 
We note here that, from analytical and numerical estimates of the covariance matrix, we find the covariance between radial bins negligible on the scales of interest here.

Galaxy-galaxy lensing offers a indirect measure of the projected mass density:  
\begin{equation}
\Delta \Sigma(R) \equiv \bar{\Sigma}(<R) - \Sigma(R),
\label{Eq:ESD_gamma_t}
\end{equation}
where \ESD is the difference between the surface density averaged within $R$, $\bar{\Sigma}(<R)$, and measured at $R$, $\Sigma(R)$.

\subsection{The lens sample}
\label{Sec:vol_lim_gama}

In this work, we make use of the group catalogue of the GAMA survey \citep[G3Cv7, ][]{Robotham11} and version 16 of the stellar mass\footnote{We note that stellar masses of GAMA galaxies have been estimated in \cite{Taylor11}. In short, stellar population synthesis models from Bruzual \& Charlot (2003) that assume a \citet{Chabrier03} Initial Mass Function (IMF) are fit to the {\it ugriz}-photometry from SDSS. NIR photometry from VIKING is used when the rest-frame wavelength is less than $11\,000 \mathrm{\AA}$.} catalogue, which contains approximately $180\,000$ objects, divided into three separate $12\times5$ square degrees patches \citep[]{Liske15} that completely overlap with the northern stripe of KiDS.

The G3Cv7 group catalogue is based on a friends-of-friends (FoF) algorithm, which links galaxies based on their projected and line-of-sight distance. Groups are therefore identified using spatial and spectroscopic redshift information \citep[][]{Robotham11}. The linking length used by the group finder has been calibrated using mock data \citep{Robotham11, Merson13} from the Millennium  dark matter simulation \citep{Springel05} populated using the semi-analytical model of galaxy formation described in \citet[][]{Bower06}. 
The FoF algorithm used in the Millennium simulation employs a particle linking length of $b=0.2$ times the mean interparticle distance\footnote{The EAGLE simulation catalogue used 
throughout this paper uses the same value of the linking length (see \S\ref{Sec:FOF_sim_ESD}).}.
The GAMA group catalogue has been tested against mock data and ensures reliable central-satellite distinction against interlopers for groups with 5 or more members 
($N_{\rm FoF} \geq 5$) above the completeness limit of GAMA of approximately $\newl(M_{\rm star} / \Msun)= 8$ \citep{Robotham11}.
Throughout the paper the galaxy-galaxy lensing signal is only computed for galaxies in groups with 5 or more members.

GAMA is a flux-limited survey. This results in an increasingly higher minimum luminosity or stellar mass at higher redshifts. 
This selection function can in principle be mimicked starting from a simulation box and constructing a GAMA light cone. 
Alternatively, one could restrict the observational analysis to those GAMA galaxies that would be present in a volume (rather than flux) limited sample 
but this approach would have the shortcoming that a large number of lenses would be discarded and the resulting \ESD profiles would have a significantly lower signal-to-noise ratio. We opt for the construction of a (nearly) volume-limited lens sample following the iterative methodology described in Lange et al. (2015). From this sample, we select only galaxies that reside in groups with at least 5 members.

\begin{figure} 
\begin{center}
\includegraphics[width=\columnwidth]{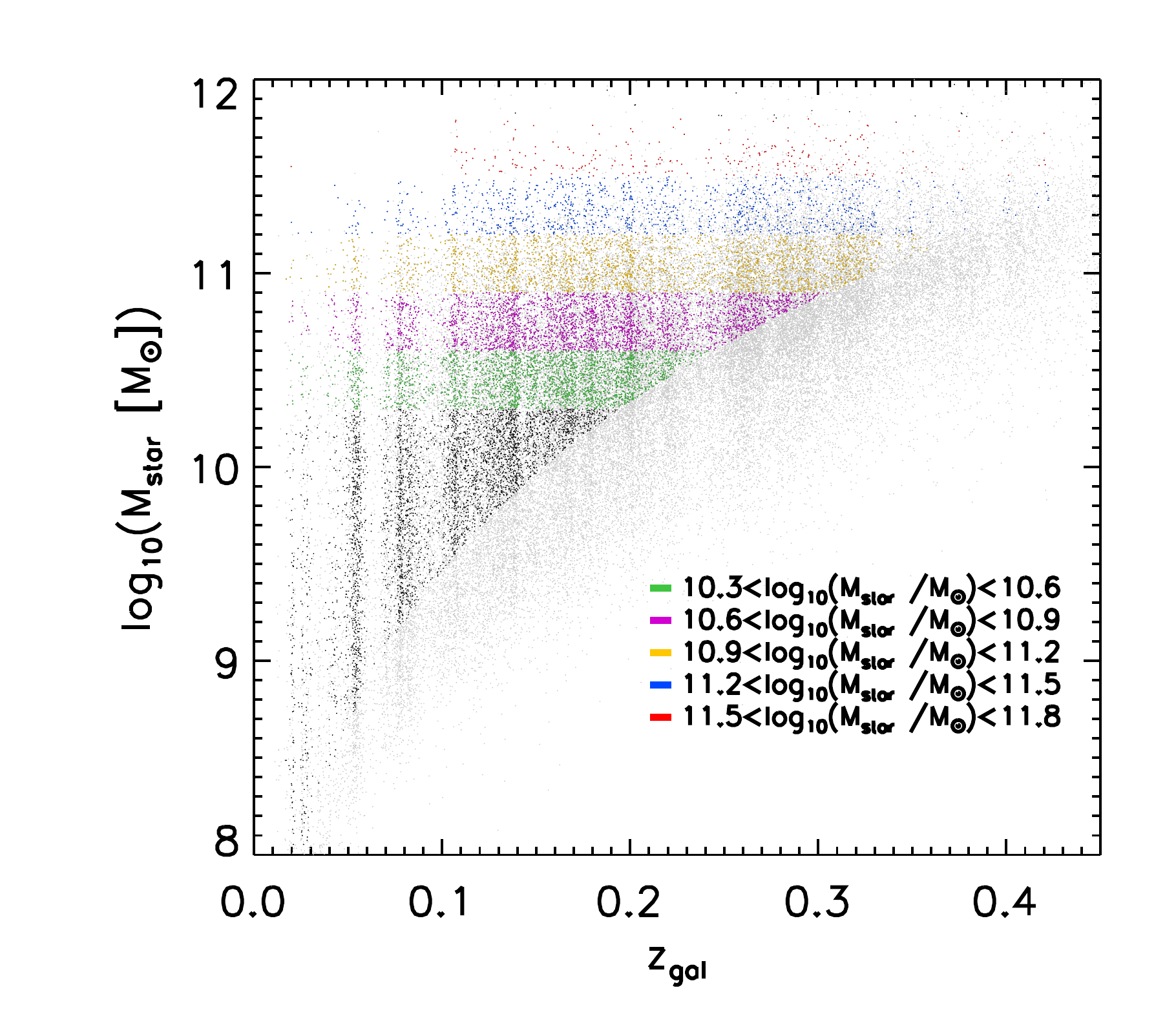}  
\end{center}
\caption{Stellar mass {\it versus} redshift of galaxies in the GAMA survey. The full sample is shown in grey. Coloured points refer to GAMA galaxies in the (nearly) volume-limited sample (see \S2.2) and in groups with at least five members.}
\label{figs:GAMA_selection}
\end{figure}

Fig.~\ref{figs:GAMA_selection} shows the stellar mass-redshift plane for the GAMA galaxies (grey points) in the area overlapping with KiDS.
Black and coloured points show which of those GAMA galaxies are in the (nearly) volume limited sample and at the same time belong to groups with five or more members.
Points are coloured according to the stellar mass bin they belong to (see column~1 of Table~1).

\section{Simulations}
\label{Sec:Eagle}
We compare the observed ESD profile to the predictions from the hydrodynamical cosmological simulations from the EAGLE project \citep{Schaye14, Crain15} with a cubic volume of $100^3\Mpc^3$. EAGLE was run using a modified version of the $N$-Body Tree-PM smoothed particle hydrodynamics (SPH) code \textsc{gadget-3}, which was last described in \citet{Springel05_gadget}. The main modifications with respect to \textsc{gadget-3} regard the formulation of the hydrodynamics, the time stepping and the subgrid physics.
Dark matter and baryons are represented by $2\times 1504^3$ particles, with an initial particle mass of $m_{\rm b}= 1.2 \times 10^6 \Msun $ and $m_{\rm dm} =  9.75  \times 10^6 M_{\odot}$ for baryons and dark matter, respectively. EAGLE was run using the set of cosmological values suggested by the initial results from the Planck mission \{$\Omega_{ \rm m}$,\, $\Omega_{ \rm b}$,\,$\sigma_{ \rm 8}$,\, $n_{ \rm s}$,\, $h$\} = \{0.307, 0.04825, 0.8288, 0.9611, 0.6777\} (Table 9; \citealt{Planck13}).

EAGLE includes element-by-element radiative cooling (for 11 elements; \citealt{Wiersma09a}), pressure and metallicity-dependent star formation \citep[][]{Schaye04, Schaye08}, with a \
\citet{Chabrier03} Initial Mass Function, stellar mass loss \citep{Wiersma09b}, thermal energy feedback from star formation \citep{DallaVecchia12}, angular momentum dependent gas accretion onto supermassive black holes \citep{Rosas13} and AGN feedback \citep{Booth09, Schaye14}.
The subgrid feedback parameters were calibrated to reproduce the present day observed galaxy stellar mass function (GSMF) as well as the observed distribution of galaxy sizes \citep{Schaye14}.
More information regarding the technical implementation of hydrodynamical aspects as well as subgrid physics can be found in \citet{Schaye14}.

\subsection{Halo catalogue}
\label{Sec:FOF_sim_ESD}
Groups of connected particles are identified by applying the FoF algorithm to the dark matter particles using a linking length of $0.2$ times the mean inter-particle separation \citep{Davis85}. Baryons are then linked to their closest dark matter particle and they are assigned to the same FoF group, if any.
Subhaloes in the FoF group are identified using \textsc{subfind} \citep{Springel01_subfind,Dolag09}. \textsc{subfind} identifies local minima in the gravitational potential using saddle points. All particles that are gravitationally bound to a local minimum are grouped into a subhalo.
Particles that are bound to a subhalo belong to that subhalo only. We define the  subhalo centre as the position of the particle for which the gravitational potential is minimum. The mass of a subhalo is the sum of the masses of all the particles that belong to that subhalo.
The most massive subhalo is the \emph{central} subhalo of a given FoF group and all other subhaloes are \emph{satellites}.

The mass $M_{200}^{\rm crit}$ and the radius $r_{200}^{\rm crit}$ of the halo are assigned using a spherical over-density algorithm centred on the minimum of the gravitational potential, such that $r_{200}^{\rm crit}$ encompasses a region within which the mean density is 200 times the critical density of the universe. 

The group finder of EAGLE links particles in real space whereas the GAMA group finder connects members in redshift space. This difference could be particularly important if a large fraction of interlopers were wrongly assigned to groups for GAMA. However, the GAMA group finder was tested against mock catalogues and found to be robust against interlopers for groups with five or more members \citep{Robotham11}. We defer a more detailed study of the impact of adopting exactly the same grouping algorithm to a forthcoming publication by the KiDS collaboration. 

\begin{figure}
\begin{center} 
\includegraphics[width=\columnwidth]{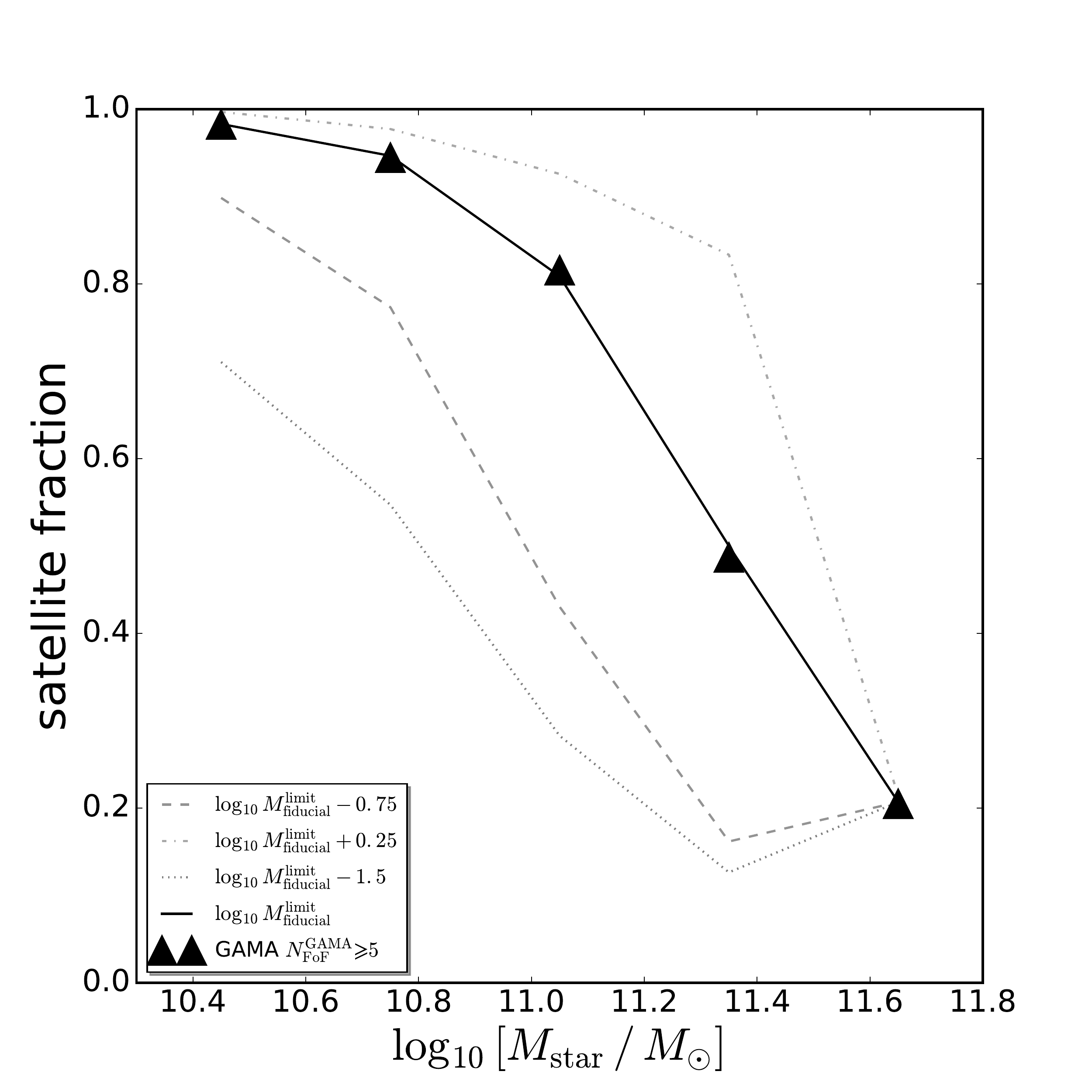} 
\end{center}
\caption{Satellite fraction, $f_{\rm sat}$, in EAGLE (black curve) obtained with a choice of $M_{\rm star}^{\rm limit}$ that reproduces the GAMA satellite fraction (black triangles). Curves with different line styles and shades of grey show the satellite fraction with a choice of the $M_{\rm star}^{\rm limit}$ of respectively -1.5, -0.75 below and +0.25 dex above the reference values.
}
\label{figs:f_sat}
\end{figure}

\subsection{Computation of the galaxy-galaxy lensing signal in EAGLE}
\label{Sec:ESD_sim}
The galaxy-galaxy lensing signal from observations measures the \ESD profile\footnote{Although this is strictly true only to the extent to which one knows the 
source redshift distribution.} (Sec.~\ref{Sec:ESD_data}). Therefore, in order to compare to the observations, we calculate the \ESD profiles from EAGLE. 
To do so, we project all the particles within a sphere with radius $2.95$ Mpc centred on the location of the subhalo onto the $x-y$ plane\footnote{We tested that the results do not differ 
significantly by choosing different projections or averaging over the three of them.}. We divide the projected radial range into 150 bins equally spaced in log-space. At every projected radius 
$R$, we calculate the surface density within $R$, $ \bar{\Sigma}(<R)$, as the sum of the mass of all the particles within the projected radius $R$, $M(<R)$, divided by the area $A=\pi R^2$. 
The surface density at $R$, $\Sigma(R)$, is the mass enclosed in the annulus with inner radius $(R-\delta R/2)$ and outer radius $(R+\delta R/2)$ divided by the area $2\pi R \delta R$, 
where $\delta \newl R = \newl(2.95[$Mpc$])/150$. 
We tested different choices for the shape and extent of the projection volume, as in principle, the lensing signal is affected by all the matter between the source and the lens and not only 
that residing within a certain distance from the lens. We verified that projecting a cylindrical section around the centre of a subhalo instead of a sphere has a negligible  effect on the ESD 
profile at all scales of interest in this paper (from virtually null at $R<0.7$ Mpc to a few percent at $R\approx2$ Mpc) but a large impact on the computation time. We thus opted for spherical 
regions. We also tested the impact of using different radii. Specifically, we found that using spheres of $4.43$ Mpc instead of $2.95$ Mpc has a negligible effect on the signal.

Subhaloes are binned according to their stellar mass, calculated as the sum over all stellar particles that belong to the subhalo. The $\Delta \Sigma$ in a given stellar mass bin is then calculated by averaging the $\Delta \Sigma$ profiles of single subhaloes.
The statistical errors are calculated using bootstrapping: galaxies in each mass bin are re-sampled 1000 times and the range of values that count for the 95\% of the distribution is taken as the 2-sigma error for the ESD profiles from the simulation.

\subsection{Selection function}
\label{Sec:sel_func_eagle}
In order to avoid selection bias, it is important that the sample of galaxies that is selected in the simulations is a fair representation of the galaxy sample in GAMA. 
The GAMA galaxy sample (nearly volume-limited and with groups with 5 or more members) has a median redshift of $z=0.16$ and hence we compare 
the corresponding galaxy-galaxy lensing signals with those obtained from the snapshot of the EAGLE simulation closest in redshift, i.e. $z$=0.18. 
The slight discrepancy in redshift is likely unimportant as from $z$=0.25 to $z$=0 there is little evolution in the GSMF \citep{Furlong15}. 
We verified that the effect of using EAGLE galaxies at $z$=0 is indeed negligible.

A robust discrimination between satellites and central galaxies is obtained by restricting our sample to galaxies that belong to groups with at least five members. 
To mimic this selection, we need to impose a minimum stellar mass from which we start counting group members in EAGLE. 
The choice of this $M_{\rm star}^{\rm limit}$ is somewhat arbitrary and  could alter the ratio between the number of satellite and central galaxies in a given stellar mass bin. 
By increasing the stellar mass limit, the number of central galaxies in groups that have four or more satellites above the mass limit is reduced, 
whereas the number of satellites of a given stellar mass remains mostly unchanged. Therefore, increasing the stellar mass limit has the net effect of increasing the satellite fraction. 
We choose the value of $M_{\rm star}^{\rm limit}$ that results in the ratio of satellite to total galaxies found in GAMA for a given stellar mass bin. 
In the rest of the paper we also show the effect of a different choice of $M_{\rm star}^{\rm limit}$ on the \ESD profile results from EAGLE. 

Fig.~\ref{figs:f_sat} shows the satellite fraction in EAGLE for different choices of $M_{\rm star}^{\rm limit}$. The black triangles show the satellite fraction in our galaxy sample from GAMA 
for galaxies in the same stellar mass bins. The black line represents the satellite fraction in EAGLE if we choose the value of $M_{\rm star}^{\rm limit}$ that reproduces the GAMA satellite 
fraction (see Table~1). With different linestyles and shades of grey we show the satellite fraction with a choice of the $M_{\rm star}^{\rm limit}$ of respectively -1.5, -0.75 below and +0.25 
dex above the values that reproduce the GAMA satellite fraction. For the fiducial choice of $M_{\rm star}^{\rm limit}$ the satellite fraction of GAMA is reproduced by construction, 
but we note that this would not necessarily be the case if the number of galaxies in a stellar mass bin were too small to recover the exact satellite fraction.
Decreasing (increasing) the value of $M_{\rm star}^{\rm limit}$ with respect to the fiducial value, has the net effect of decreasing (increasing) the satellite fraction. 
The fiducial values of $M_{\rm star}^{\rm limit}$ in each stellar mass bin are $(9.46, 9.91, 9.96, 10.33, --)$, see also column $(11)$ of Table~1. We note that the value of $M_{\rm star}^{\rm limit}$ is ill-defined for the most massive bin. In fact, the haloes that enter in this bin satisfy the richness cut for every value of $M_{\rm star}^{\rm limit}$ that is lower than the lower limit of the bin itself ($\log[M_{\rm star}/M_{\odot}] = 11.5$). We also note that the fiducial values of $M_{\rm star}^{\rm limit}$ are close to the completeness limit at z=0.18 of the specific GAMA galaxy group sample adopted throughout the paper.

Since the value of the satellite fraction, in our approximation of the GAMA selection function, is essential for the calculation of the combined signals from satellite and central galaxies, the choice of $M_{\rm star}^{\rm limit}$ has a major effect on the comparison with observations when galaxies are not separated in centrals and satellites (see \S4.3).

\section{Results}
\label{Sec:Result_ESD}
In the following we present the results for the excess surface density \ESD computed from the simulations (for details see \S\ref{Sec:ESD_sim}). 
We divide galaxies into five stellar mass bins ranging from $\newl(M_{\rm star}/\Msun)$ $= 10.3$ to $\newl(M_{\rm star}/\Msun)$ $= 11.8$.
In the simulations we consider all stellar mass particles bound to a subhalo for the stellar mass determination. 
We note that this choice may overestimate the stellar mass content since in observations stars in galaxy outskirts are often not detectable. 
We address this caveat by correcting the stellar mass of GAMA galaxies by a multiplicative factor given by the ratio between the galaxy's measured flux in the r band and the integral 
of its Sersic profile up to infinity \citep{Taylor11}. In this way we correct the stellar mass of galaxies by taking into account their undetected flux. 
An alternative approach would be to consider only stellar particles within a 30 kpc aperture for the stellar mass calculation in EAGLE \citep[see the discussion in][]{Schaye14}. 
Similarly, we would need to correct the observed stellar mass by the multiplicative factor given by the ratio between the measured flux (r band) of the galaxy and its integrated 
Sersic profile up to 30 kpc. We tested this alternative approach, leading to very similar results with the disadvantage of reducing the number of galaxies available from the EAGLE 
simulations in the highest stellar mass bins. We therefore opted for the former approach. The ESD in a given stellar mass bin is computed by stacking the \ESD of all galaxies 
in that mass bin. 

We compare each prediction from the simulation to the corresponding data from KiDSxGAMA. We first present results for central and satellite galaxies separately (see \S\ref{Sec:ESD_sim_cen} and \S\ref{Sec:ESD_sim_sat}). We then present the results for both galaxy types combined (\S\ref{Sec:ESD_sim_tot}). This signal is the linear combination of the signal from satellite and central galaxies, where the relative importance of the two terms is modulated by the value of the satellite fraction (\S\ref{Sec:ESD_data_comparison_tot}).

\begin{figure} 
\begin{center}
\includegraphics[width=\columnwidth]{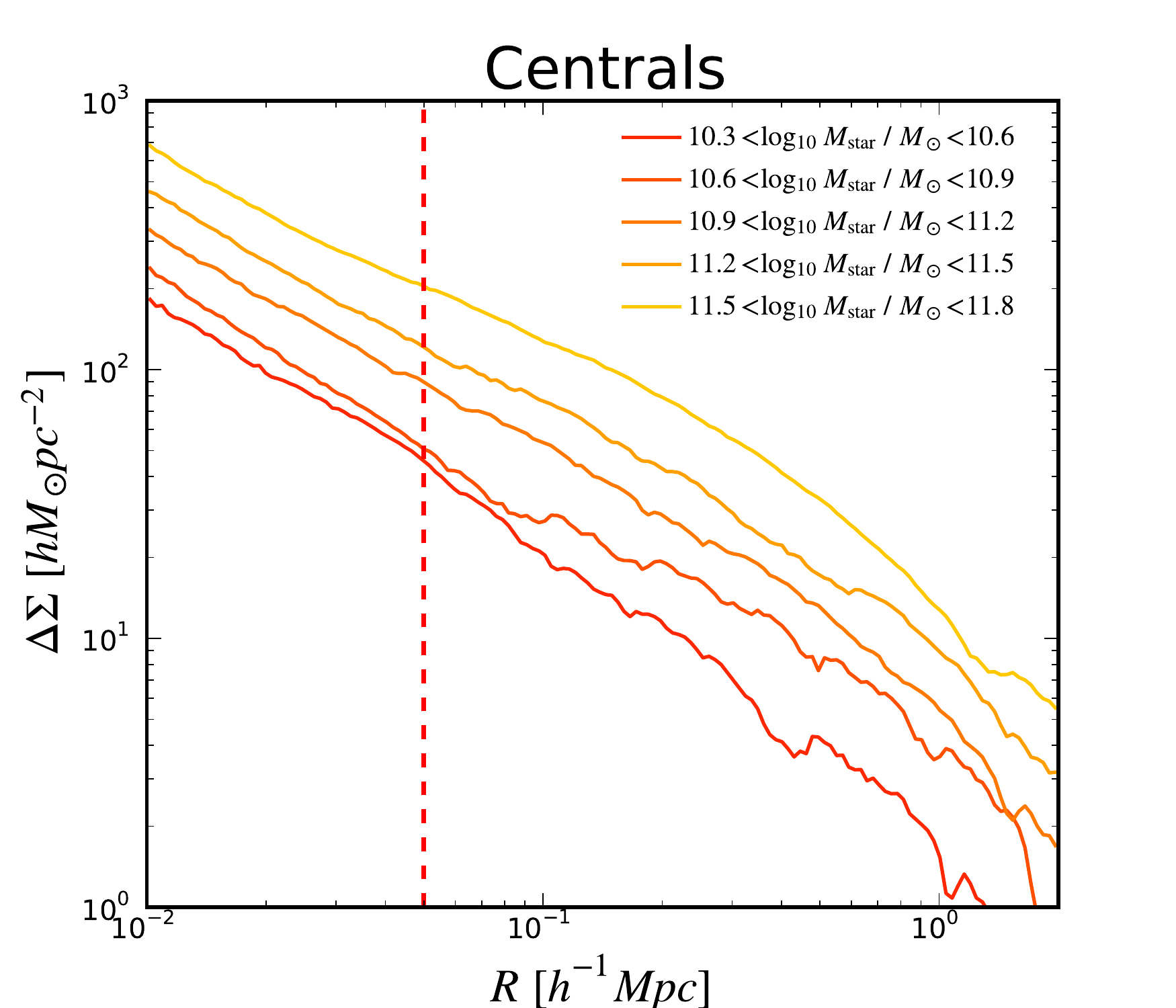}  
\end{center}
\caption{Profiles of the excess surface density, \ESD, of matter around central galaxies up to projected separations of $2\hMpc$ from the centre of the galaxy. To mimic the GAMA selection function, only galaxies hosted by groups with five or more members with masses above the stellar mass limit listed in column 11 of Table 1 are taken into account for this analysis. Central galaxies are divided into five stellar mass bins ranging from $\newl(M_{\rm star}/ \Msun)= 10.3$ to $\newl(M_{\rm star}/\Msun)= 11.8$. The vertical dashed line marks $R = 0.05 \hMpc$ representative of the scales at which the inner part of the dark matter halo dominates the signal.
}
\label{figs:ESD_multibin_cen}
\end{figure}

\begin{figure*} 
\begin{center} 
\includegraphics[width=2.0\columnwidth]{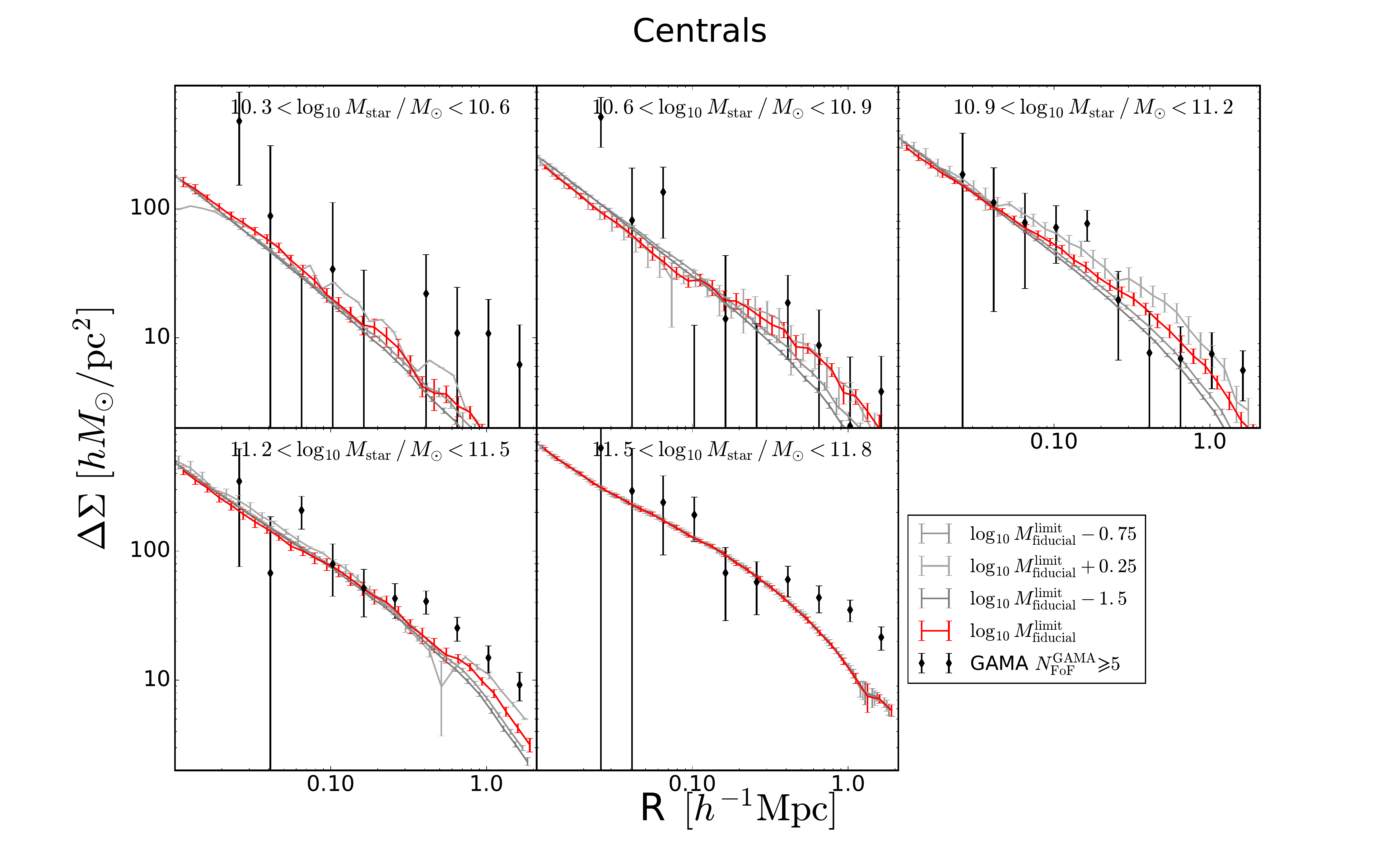}
\end{center}
\caption{Excess surface density profiles from KiDSxGAMA (black diamonds) and in the EAGLE simulation (red curves) for central galaxies hosted by groups with 5 or more members 
that each have stellar masses greater than $M_{\rm star}^{\rm limit}$ (listed in column 11 of Table 1) in order to mimic the GAMA selection of galaxies.  
Each panel contains a different bin in central galaxy stellar mass. Asymmetric error bars show the 2-$\sigma$ error in each $R$ bin. 
Curves with different shades of grey show the ESD profiles in EAGLE with a choice of the $M_{\rm star}^{\rm limit}$ of respectively -1.5, -0.75 dex below and +0.25 dex above the values that reproduce the GAMA satellite fraction.
}
\label{figs:EAGLEvsKIDS_cen}
\end{figure*}

\subsection{The galaxy-galaxy lensing signal around central galaxies}
\label{Sec:ESD_sim_cen}
Fig.~\ref{figs:ESD_multibin_cen} shows the ESD profile around central galaxies in the EAGLE simulation as a function of the projected distance from the centre of the galaxy. 
For all mass bins \ESD is a decreasing function of the projected radius. Fluctuations in the excess surface density profiles can arise due to the presence of matter 
associated to satellite galaxies, but these are usually not massive enough to significantly alter the azimuthally averaged excess surface density profile. 
Moreover, since the signal is averaged over many galaxies, any deviation due to the presence of a relatively massive satellite would be averaged out in the stacking process.

Table \ref{tbl_ch5:stat_ESD} reports values of the mean subhalo mass $M^{\rm cen}_{\rm sub}$ for each stellar mass bin.
The $\Delta\Sigma(R= 0.05 \hMpc)$ (the intersection between the red dashed line in Fig.~\ref{figs:ESD_multibin_cen} 
and the \ESD profiles for different stellar mass bins) and the mean mass $M^{\rm cen}_{\rm sub}$ are monotonically increasing functions of the stellar mass. 
Both $\Delta\Sigma(R= 0.05 \hMpc)$ and $M^{\rm cen}_{\rm sub}$ are  approximated reasonably well by single power law functions of the stellar mass (not shown here), 
albeit with different coefficients. $\Delta\Sigma(R= 0.05 \hMpc)$ shows a weaker dependence on stellar mass with respect to $M_{\rm sub}$ which, in this stellar mass range, 
has a power law coefficient close to unity. Central galaxies with higher \ESD amplitudes are hosted by more massive haloes.  
Therefore, as expected, the amplitude of the \ESD profile at small scales is a proxy for the typical mass of the subhaloes hosting central galaxies in a given stellar mass bin.

\subsubsection{Comparison with observations}
\label{Sec:ESD_data_comparison_cen}
Fig.~\ref{figs:EAGLEvsKIDS_cen} shows  the \ESD signal in EAGLE (red curves) whereas \ESD from the observations is indicated with black diamonds and vertical error bars. 
Curves with different shades of grey show the ESD profiles in EAGLE with a different choice of the $M_{\rm star}^{\rm limit}$ (see \S \ref{Sec:sel_func_eagle}). 
For stellar masses $10.3<\newl(M_{\rm star}/\Msun)< 10.6$, the uncertainties in the data are large due to the limited number of low-mass  galaxies that are centrals in rich groups 
($N_{\rm FoF}^{\rm GAMA}\geq5$) and therefore are not representative of the entire central galaxy population \citep[][]{Viola15}. 
For stellar masses $10.6<\newl(M_{\rm star}/\Msun)< 11.8$ the uncertainties on the measurements are smaller and the radial dependence of the signal is better constrained. 
We find an overall agreement between data and predictions from the simulation and in what follows we discuss some features in more detail.

The agreement between the ESD in EAGLE and KiDS suggests that central galaxies, with masses 
$10.6<\newl(M_{\rm star}/\Msun)< 11.5$ in the simulation are hosted  by subhaloes of approximately the correct mass and the right density profile.
This is perhaps not surprising considering that EAGLE was calibrated to broadly reproduce the GSMF (mostly composed by central galaxies) and therefore to assign approximately the correct stellar mass to subhaloes. For $11.2<\newl(M_{\rm star}/\Msun)< 11.8$ the observed \ESD seems to favour a shallower excess surface density profile at radii larger than $400\hkpc$. This might reflect a box-size effect, as more massive (more extended and less concentrated) haloes might be missing in the small volume probed by the EAGLE simulation.

The mean host halo masses predicted by EAGLE for galaxies in the five stellar mass bins shown can be found in Table \ref{tbl_ch5:stat_ESD}, column (2).

We have computed analytical \ESD profiles corresponding to haloes with \citet[][hereafter NFW]{NFW} matter density profiles for the halo masses reported in column (2) of Table \ref{tbl_ch5:stat_ESD}. These analytical profiles reproduce the overall normalisation of the signal but poorly match the radial dependence of the numerical profiles. In Appendix A, we discuss this test in detail, and we also comment on the cause of the limitations of simple analytical model in accurately describing the \ESD profiles obtained from simulations.

In the case of central galaxies the choice of $M_{\rm star}^{\rm limit}$ has a small effect on the ESD profile computed from the simulations as can be seen by comparing the grey lines in Fig.~\ref{figs:EAGLEvsKIDS_cen}. To quantify this, we employ the following statistics:
\begin{equation}
\chi^2_{\rm red}= \frac{1}{(N_{\rm points}-1)} \sum_{\rm i} \frac{  (\Delta\Sigma^{\rm EAGLE}_i - \Delta\Sigma^{\rm data}_i)^2  }{{\sigma_i^{\rm EAGLE}}^2+{\sigma_i^{\rm data}}^2  },
\label{Eq:chi_sq}
\end{equation}
where $N_{\rm points}$ is the number of stellar mass bins times the number of data points per bin and $i$ is an index running through all 60 data points. We obtain values $\chi^2_{\rm red}=1.4$ for the fiducial value of $M_{\rm star}^{\rm limit}$. We note that four points in each of the two most massive stellar mass bins lead to most of the deviations of $\chi^2_{\rm red}$ from unity.  Furthermore, $\chi^2_{\rm red}$ ranges from 1.4 to 1.8 as we change $M_{\rm star}^{\rm limit}$ from its fiducial value to the fiducial -1.5.  We note that, throughout the paper, we are neglecting any off-diagonal terms of the covariance matrix. Although this might have a (supposedly small) effect on the absolute value of the $\chi^2_{\rm red}$, we are here mostly concerned with relative differences among models with different choices of a limiting stellar mass. In the context of this comparison, we consider the differences reported above not worth further investigations.

Higher values of $M_{\rm star}^{\rm limit}$ tend to produce higher amplitudes of the ESD profiles since higher mass subhaloes are being selected.
The relative insensitivity on the exact choice of $M_{\rm star}^{\rm limit}$ suggest that for a comparison of ESD profiles of central galaxies only, the exact details of the galaxy selection are not crucial. We anticipate that the same argument is not applicable when the ESD profiles of central and satellite galaxies are analysed jointly since the choice of $M_{\rm star}^{\rm limit}$ determines the satellite fraction which in turn plays a major role in establishing how the ESD profiles of central and satellite galaxies are combined (see \S\ref{Sec:ESD_sim_tot}).

\subsection{The galaxy-galaxy lensing signal around satellite galaxies}
\label{Sec:ESD_sim_sat}

Unlike central galaxies, the \ESD profiles of the satellites galaxies are \emph{not} necessarily expected to be simply decreasing functions of the separation from the centre.
For a single satellite galaxy the profile should become negative at the projected separation from the centre of the host halo \citep{Yang06,Sifon15}. 
This effect is due to the surface density at the centre of the host halo being larger than the mean internal surface density, 
$\Sigma(R_{\rm  centre}^{\rm halo}) > \bar{\Sigma}(<R_{\rm centre}^{\rm halo})$.
At larger separations than the separation to the host halo, 
the \ESD profile first increases due to the inclusion of the centre of the host halo in the term $\bar{\Sigma}(<R)$, before decreasing again at still larger separations.
Stacking the \ESD of satellites in a given stellar mass bin smooths out the negative parts of the profiles since the separations between satellites and their host halo vary. 
However, the increase in the signal at larger radii is preserved by the stacking. 

\begin{figure} 
\begin{center} 
\includegraphics[width=\columnwidth]{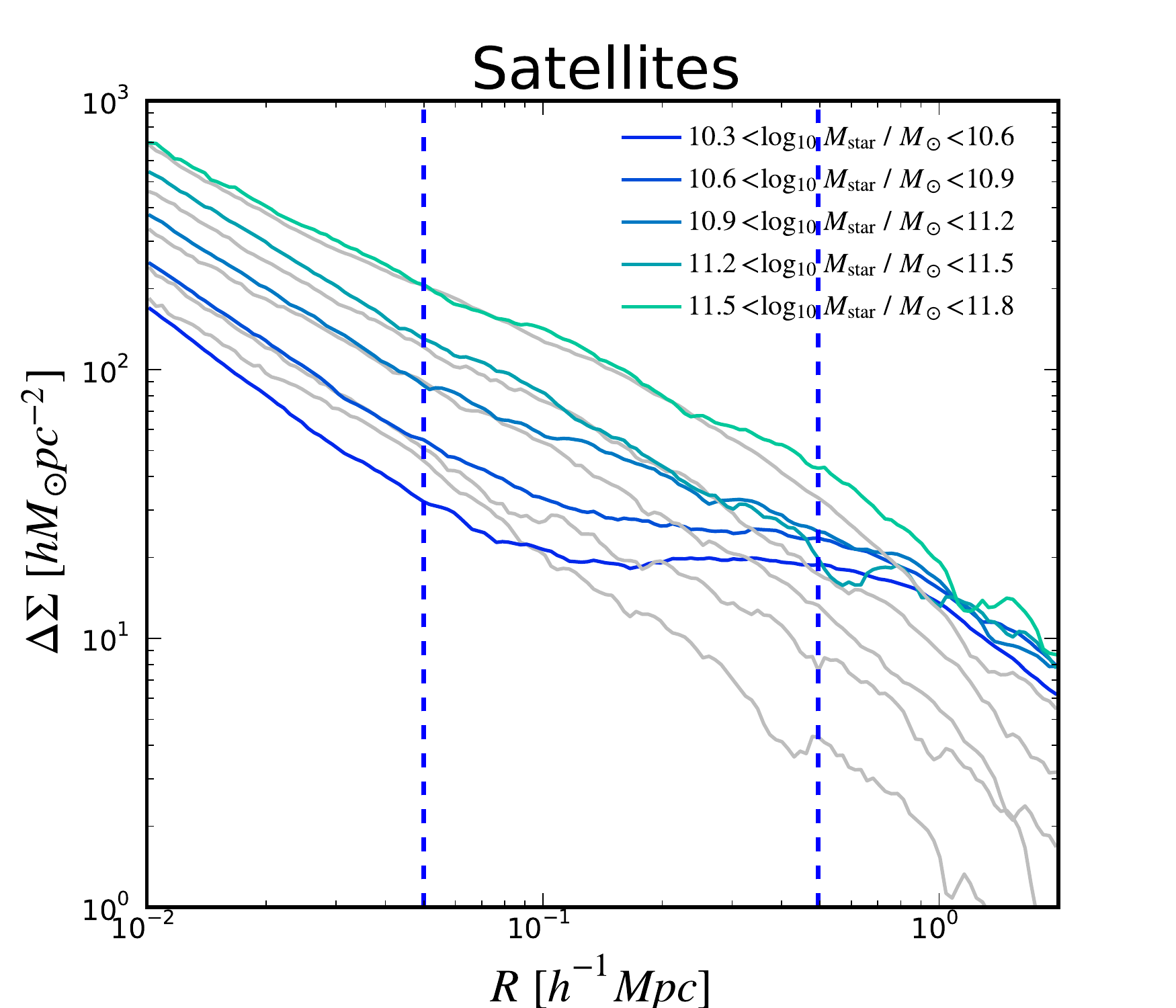}
\end{center}
\caption{
As Fig.~\ref{figs:ESD_multibin_cen} but for satellite galaxies. To ease the comparison the results for the central galaxies are reported with grey curves. 
The two vertical lines mark $R= 0.05 \hMpc$ and the $R = 0.5 \hMpc$.
}
\label{figs:ESD_multibin_sat}
\end{figure}

Fig.~\ref{figs:ESD_multibin_sat} shows the \ESD profile of satellite galaxies in the EAGLE simulation.
The amplitude of the \ESD profile at small separations ($R = 0.05 \hMpc$) is an increasing function of the stellar mass of the satellite. 
The same trend is shared by the average subhalo mass for satellite galaxies since satellites with higher stellar masses tend to be hosted 
by more massive dark matter subhaloes (see Table \ref{tbl_ch5:stat_ESD}, column 5).
As in the case of central galaxies, the similar dependence on the stellar mass suggests that the amplitude of \ESD at small separations can be considered a proxy for the mass 
of the subhalo hosting the satellite galaxy.

The radius at which the \ESD profile starts to be dominated by the host halo mass (the satellite bump) increases with stellar mass. This effect is driven by the change in the average 
distance between satellites and their host haloes, which increases from $\approx 880$kpc to $\approx 2800$ kpc in the mass range considered (see Table \ref{tbl_ch5:stat_ESD}, column 7).

For larger separations ($R= 0.5 \hMpc$), the \ESD profile starts to be dominated by the contribution of the halo hosting the satellite galaxy. In this case \ESD shares a similar trend with stellar mass as the mean host halo mass for satellite galaxies, $M_{200}^{\rm crit}$ (see Table \ref{tbl_ch5:stat_ESD}, column 3). 

The amplitude of the satellite bump is similar for all the stellar mass bins, which can be explained by the fact that the richness cut effectively selects host haloes by mass. Indeed, most of 
the satellites with stellar mass $10.3<\newl(M_{\rm star}/\Msun)< 11.8$ reside in host haloes of  mass $13.95<\newl[M_{200}^{\rm crit}/M_{\odot}]< 14.24$.
The prominence of the satellite bump with respect to the overall normalisation decreases with stellar mass, 
a trend that is explained by the fact that the ratio $ M_{\rm sub}^{\rm sat}/M_{200}^{\rm crit}$ increases from 0.03 to 0.3 in the considered mass range 
(see Table \ref{tbl_ch5:stat_ESD}, column 6).  

The similar dependence of \ESD with halo mass at larger radii highlights the fact that the amplitude of the satellite bump is tightly correlated to the host halo mass. In principle the amplitude of the satellite bump should depend on the satellite's subhalo mass as well as on the host halo mass. In practice the satellite's subhalo mass is, except for the highest stellar mass bin, a small fraction of the host halo mass and therefore it plays a minor role in setting the amplitude of the satellite bump. 

\subsubsection{Comparison with observations}
\label{Sec:ESD_data_comparison_sat}

\begin{figure*} 
\begin{center} 
\includegraphics[width=2.0\columnwidth]{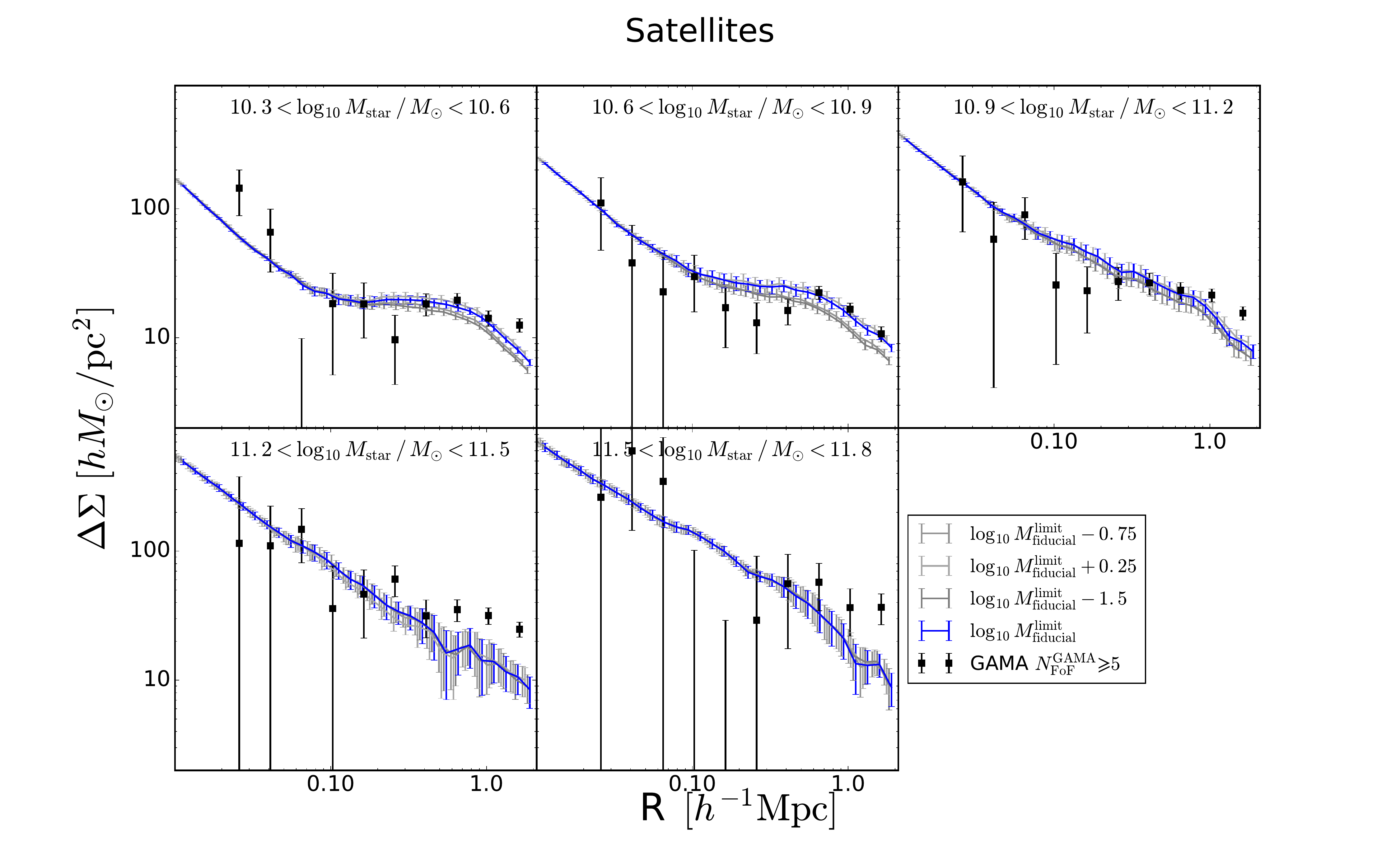}
\end{center}
\caption{
Excess surface density profiles from KiDSxGAMA (black squares) and in the EAGLE simulation (blue curves) for satellite galaxies hosted by groups with 5 or more members that each have stellar masses greater than $M_{\rm star}^{\rm limit}$ {(listed in column (11) of Table~1)} in order to mimic the GAMA selection of galaxies. Each panel contains a different bin in satellite galaxy stellar mass. As in Fig.~\ref{figs:EAGLEvsKIDS_cen} , asymmetric error bars show the 2-$\sigma$ error in each $R$ bin. Curves with different shades of grey show the ESD profiles in EAGLE with a choice of the $M_{\rm star}^{\rm limit}$ of respectively -1.5, -0.75 dex below and +0.25 dex above the values that reproduce the GAMA satellite fraction.
}
\label{figs:EAGLEvsKIDS_sat}
\end{figure*}

Fig.~\ref{figs:EAGLEvsKIDS_sat} shows the comparison between the observed \ESD profile of satellite galaxies (black squares) 
and the corresponding signal in the EAGLE simulation (blue curves) for the five stellar mass bins.
The ESD profiles computed for different choices of $M_{\rm star}^{\rm limit}$ are shown in grey. 
For $10.3<\newl(M_{\rm star}/\Msun)< 10.9$ there is an overall broad agreement between simulation predictions and observations.

For $\newl(M_{\rm star}/\Msun)>10.6$ the  normalization of the ESD profile at small ($0.03 < R < 0.2 \hMpc$) scales is higher in the simulations than in the observations although at low significance (less than two sigma).
 
For stellar masses $10.9<\newl(M_{\rm star}/\Msun)< 11.8$ the data show a higher amplitude for the satellite bump with respect to the simulations. This unreproduced feature could be explained by the fact that in EAGLE, due to its relative small volume, massive clusters and the satellite galaxies that they host are underrepresented. The inclusion of those satellites would increase the amplitude of the satellite bump which depends strongly on the host halo mass (see previous section). Indeed, by analysing a version of EAGLE that has the same mass resolution but an eighth of the volume, we find that the amplitude of the satellite bump decreases, an effect that is more important at higher stellar masses.

As was also seen for the ESD profile of central galaxies, the choice of $M_{\rm star}^{\rm limit}$ has only a relatively minor effect on the ESD profile of satellite galaxies as computed from the simulation. In fact, the reduced $\chi^2$ between the model and the data increases from its fiducial value of $\chi^2_{\rm red}=2.7$   to $\chi^2_{\rm red}=3.7$   in the case of $\newl (M_{\rm star}^{\rm limit})=-1.5$. 

\subsection{The galaxy-galaxy lensing signal around all galaxies}
\label{Sec:ESD_sim_tot}

In this section we present the ESD calculated considering all galaxies without distinguishing between centrals and satellites (studying only galaxies in rich groups). 
The \ESD profile of the whole population of galaxies of a given stellar mass is a linear combination of the profiles for satellites, $\Delta\Sigma_{\rm sat}$, 
and centrals, $\Delta\Sigma_{\rm cen}$:
\begin{equation}
\Delta\Sigma = f_{\rm sat}\Delta\Sigma_{\rm sat} + (1-f_{\rm sat})\Delta\Sigma_{\rm cen} \, ,
\label{eq:ESD_tot}
\end{equation}
where $f_{\rm sat}$ is the satellite fraction of galaxies in a given stellar mass bin. The relative importance of each term is set by the value of $f_{\rm sat}$. 
Therefore the \ESD profile of the whole galaxy population lies in between those for satellite and central galaxies.

\subsubsection{Comparison with observations}
\label{Sec:ESD_data_comparison_tot}

\begin{figure*}
\begin{center} 
\includegraphics[width=2.0\columnwidth]{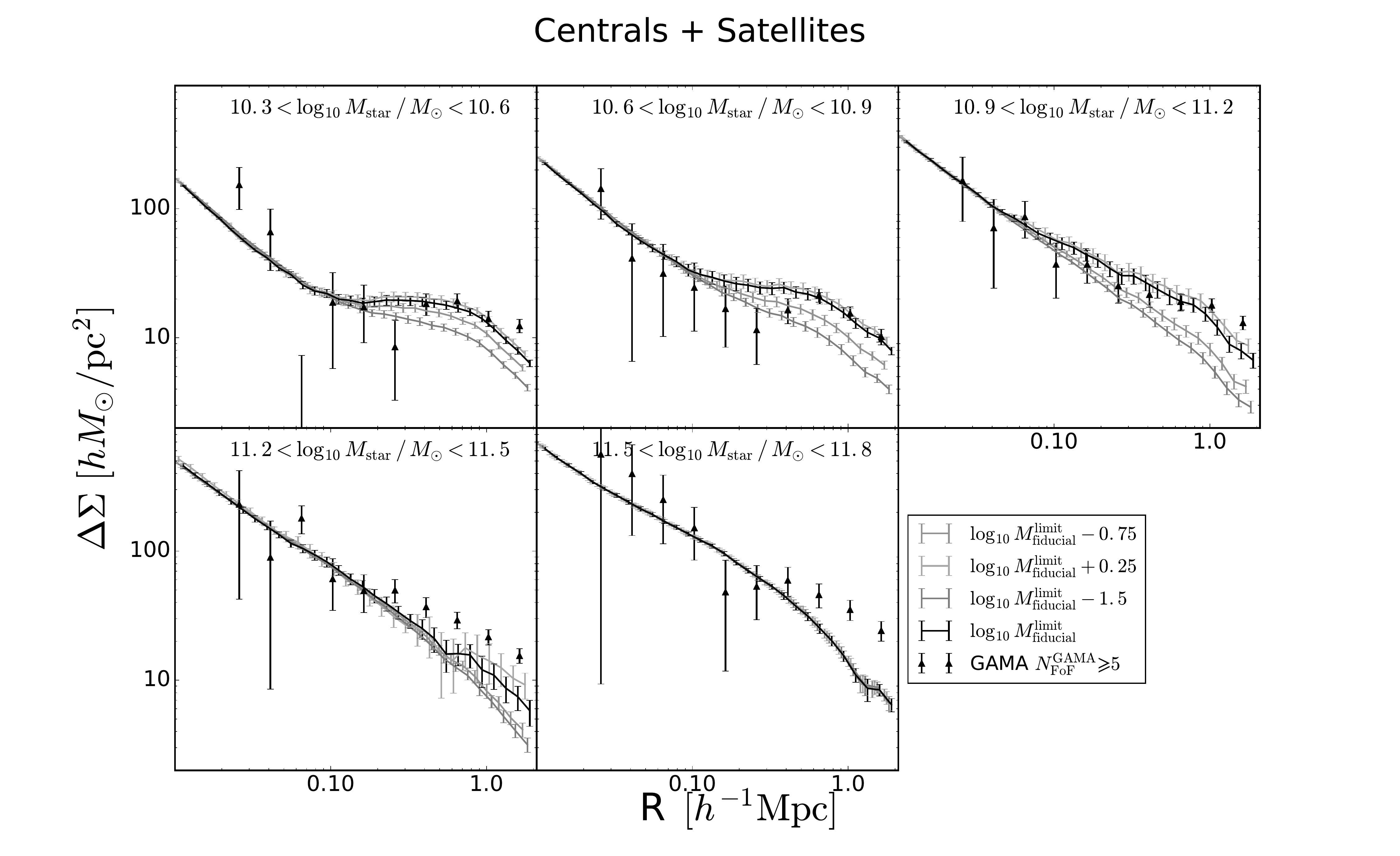}
\end{center}
\caption{
Excess surface density profiles from KiDSxGAMA (black triangles) and in the EAGLE simulation (black curves) for all galaxies hosted by groups with 5 or more members that each have stellar masses greater than $M_{\rm star}^{\rm limit}$ (listed in column (11) of Table~1) in order to mimic the GAMA selection of galaxies. Each panel refers to a different bin in galaxy stellar mass. As in Fig.~\ref{figs:EAGLEvsKIDS_cen} , asymmetric error bars show the 2-$\sigma$ error in each $R$ bin. Curves with different shades of grey show the ESD profiles in EAGLE with a choice of the $M_{\rm star}^{\rm limit}$ of respectively -1.5, -0.75 dex below and +0.25 dex above the values that reproduce the GAMA satellite fraction.
}
\label{figs:EAGLEvsKIDS_tot}
\end{figure*}

Fig.~\ref{figs:EAGLEvsKIDS_tot} shows the comparison of the \ESD profiles  obtained from observations (black triangles) and the EAGLE simulation (black curves). 
As in previous figures the \ESD profiles for different choices of $M_{\rm star}^{\rm limit}$ are shown in grey.

Most of the differences between \ESD in the simulation and observations are in line with what we expect from our previous results. 
Specifically, the amplitude of the satellite bump at $11.2<\newl(M_{\rm star}/\Msun)< 11.8$ arises from the same feature already present in the \ESD profile of satellite and central galaxies, as \ESD for all galaxies is a linear combination of the two (see Eq.~\ref{eq:ESD_tot}).

The degree of agreement between the EAGLE and GAMA results is driven by the choice of a $M_{\rm star}^{\rm limit}$ that reproduces the satellite fraction of GAMA. In fact, for different choices of $M_{\rm star}^{\rm limit}$, the agreement between the simulation and observations worsens considerably. The $\chi^2_{\rm red}$ between the model and the data increases from its fiducial value of $\chi^2_{\rm red}=2.7$   to $\chi^2_{\rm red}=8$   in the case of $\newl (M_{\rm star}^{\rm limit})=-1.5$. This dependence of $\chi^2_{\rm red}$ on the choice of $M_{\rm star}^{\rm limit}$ is considerably stronger than when satellites and centrals are analysed separately, suggesting that particular care has to be taken when selecting groups in EAGLE when satellites and centrals are analysed jointly. On the other hand, this analysis shows that the GGL signal has the potential to test the mix of satellites and centrals predicted by simulations. 

\section{Possible limitations of the comparison}
\label{Sec:discussion}

In this section we discuss some of the limitations of our study and highlight possible future improvements. The main issues are:
\begin{itemize}
\item In the comparison between simulation and observations an important role is played by stellar mass errors, both random and systematic.
We consider here the effect of a random error of $\sim 0.1$ dex \citep[][]{Behroozi13} associated with random uncertainties in the stellar mass estimation from broadband photometry. 
We are not considering the effect of systematic errors that might arise  from different choices in the stellar population synthesis model or in the initial stellar mass function\footnote{These 
errors can be significantly larger \citep[$\sim 0.3$ - 0.4 dex, see][]{Conroy09, Behroozi10, Pforr12, Mitchell13} than the random error considered here. However, as described in \S~2.2 and 
\S~3, both data and simulation assume a \citet{Chabrier03} Initial Mass Function}.  Since the number density of galaxies decreases with stellar mass, random errors always scatter more 
low-mass galaxies  to high masses than viceversa.
The importance of this effect depends on the steepness of the galaxy stellar mass function (GSMF). For low masses, $\newl(M_{\rm star}/\Msun)< 10.9$,
where the GSMF is reasonably flat, a comparable number of galaxies is scattered towards higher and lower masses. On the other hand, for higher stellar masses where the GSMF is 
steeper, relatively more low-mass galaxies are scattered towards higher masses. Therefore the effect of random errors is expected to be stronger at higher masses \citep[e.g.][]{Furlong15} .
We verified that the uncertainties in the stellar mass determinations play a very minor role for all stellar mass bins. The effect of random errors on the \ESD profiles is well within the errors 
on the simulation results.

\item The group finder of EAGLE identifies groups in real space whereas the GAMA group finder uses redshift space. This may cause differences in the \ESD profile, in particular if interlopers are wrongly assigned to groups, which would artificially increase the richness of the observed group. Therefore, the observed signal would be artificially lowered by the contribution of less massive groups hosting fewer than five members. 
To be fully consistent, the same algorithm should be employed in both simulations and observations. 

\item The centring in observations is done according to the light emitted by the galaxies -- the centre of a group is defined as the location of the Brightest Group Galaxy -- whereas in simulations we adopt the position of the particle with the minimum gravitational potential as the centre. \citet{Schaller15}, have shown that in EAGLE the majority of the galaxies ($>95\%$) have offsets between the centre of mass of their stellar and dark matter distribution that are smaller than the simulation's gravitational softening length ($\sim700 {\rm pc}$). Therefore, this effect is unlikely to be important. It should be mentioned though that the galaxy residing at the centre of the host halo is not necessarily the brightest. A more detailed comparison would then require to employ the same definition of centre in both data and simulations. 

\item In this work we mostly assume that the good agreement between the simulation and observations stems from the ability of EAGLE to reproduce the observed GSMF. A comprehensive study should be made to test how sensitive this agreement is to the level at which the GSMF is reproduced by the simulations. This can be studied by employing the EAGLE models using different subgrid parameters \citep[][]{Crain15}, although these simulations use volumes that are at least a factor of eight smaller than the main EAGLE run, which may be problematic. 

\end{itemize}

\section{Conclusions}
\label{Sec:Conclusions_ESD}

In this work we compare the excess surface density profile $\Delta \Sigma (R)$ obtained from the state-of-the-art hydrodynamical cosmological EAGLE simulation to the observed 
weak galaxy-galaxy lensing signal using (source) galaxies with accurate shape measurements from the KiDS survey around spectroscopically confirmed (lens) galaxies from the 
GAMA survey (referred throughout the paper as KiDSxGAMA). 
Results are presented for (lens) central and satellite galaxies in five logarithmically equi-spaced stellar mass bins in the range $10.3<\newl(M_{\rm star}/\Msun)< 11.8$.

The GAMA survey is 98\% complete down to {\it r}-band magnitude 19.8. This yields a relatively simple selection function. We mimic this selection function by taking  galaxies in the EAGLE 
simulation with stellar masses above $M_{\rm star}^{\rm limit}$ (about $10^{10} \Msun$, see Table 1). The precise value of this limit in stellar mass is chosen in order to reproduce the 
relative abundances of central and satellite galaxies for the different stellar mass bins in GAMA. 
To minimize the mis-identification of central and satellite galaxies, only groups with at least five members are used in the data \citep{Robotham11}. 
We apply the same `richness cut' to the simulation. 

The \ESD profile of central galaxies (Fig.~\ref{figs:ESD_multibin_cen}) in EAGLE is a decreasing function of the transverse separation with a scale-dependent logarithmic slope. 
We compare the \ESD signal of central galaxies in EAGLE with the observed signal in KiDSxGAMA. 
We find that both the normalization and the radial dependence of the signal from EAGLE are in broad agreement with the data for $\newl(M_{\rm star}/\Msun)<11.2$ (Fig.~\ref{figs:EAGLEvsKIDS_cen}).
This finding suggests that the average halo mass, as well as the projected matter density profile, around central galaxies in EAGLE is consistent with observations. The former is perhaps not surprising as the EAGLE simulation has been calibrated to reproduce the low-redshift stellar mass function and it is therefore expected to have a stellar-to-halo mass relation in agreement with observational proxies such as the stellar mass dependence of the \ESD profile. For the highest stellar mass bins ($11.2< \newl(M_{\rm star}/\Msun)<11.8$) EAGLE underestimates the signal at large radii ($\sim 1$ Mpc) most likely because its relatively modest volume ($100^3$ Mpc$^3$) leads to a lack of massive clusters.

For low stellar masses, the \ESD profile of satellite galaxies  is a non-monotonic function of the separation from the centre. This feature stems from the fact that the signal is dominated by two different components at different scales. The smallest scales ($R < 0.2 \hMpc$) are dominated by the subhalo attached to the satellite galaxy. The largest scales ($0.2< R < 2 \hMpc$ ) are dominated  by the contribution from the main host halo. For stellar masses below $\newl(M_{\rm star}/ \Msun) < 11$, the EAGLE predictions and KiDS data are in agreement at all probed scales, suggesting  that satellite galaxies in the simulations are hosted by subhaloes with the correct mass and that they reside in host haloes with the correct halo mass. The agreement is less satisfactory for galaxies with $\newl(M_{\rm star}/ \Msun) > 11.2$ for which the small volume of EAGLE plays a significant role.

When central and satellite galaxies are analysed independently, the exact choice of the galaxy selection function has a small effect on the ESD profile computed from the simulations. 
If $M_{\rm star}^{\rm limit}$ is varied by almost two orders of magnitude, the difference between the \ESD profiles is remarkably small.
However, when the ESD profiles of central and satellite galaxies are analysed jointly,
the EAGLE predictions of the ESD profile are sensitive to the selection function. 
We have calibrated the choice of the value of $M_{\rm star}^{\rm limit}$ to reproduce the GAMA satellite fraction in bins of stellar mass as this quantity encapsulates the main effect. 
However, our analysis makes apparent that, as the quality of the data improves, it will become crucial to properly mimic selection effects
to compare galaxy-galaxy lensing observations and predictions from simulations, which will enable the satellite fraction to be tested directly.

\section*{Acknowledgements}

Based on data products from observations made with ESO Telescopes at the La Silla Paranal Observatory under programme IDs 177.A-3016, 177.A-3017 and 177.A-3018, and on data products produced by Target/OmegaCEN, INAF-OACN, INAF-OAPD and the KiDS production team, on behalf of the KiDS consortium.

GAMA is a joint European-Australasian project based around a spectroscopic campaign using the Anglo-Australian Telescope. The GAMA input catalogue is based on data taken from the Sloan Digital Sky Survey and the UKIRT Infrared Deep Sky Survey. GAMA is funded by the STFC (UK), the ARC (Australia), the AAO, and the participating institutions. The GAMA website is http://www.gama-survey.org/ .

This work used the DiRAC Data Centric system at Durham University, operated by the Institute for Computational Cosmology on behalf of the STFC DiRAC HPC Facility 
(www.dirac.ac.uk). This equipment was funded by BIS National  E-infrastructure capital grant ST/K00042X/1, STFC capital grant ST/H008519/1, and STFC DiRAC 
Operations grant ST/K003267/1 and Durham University. DiRAC is part of the National E-Infrastructure. We also gratefully acknowledge PRACE for awarding us access to 
the resource Curie based in France at Tr\`es Grand Centre de Calcul. This work was sponsored by the Dutch National Computing Facilities
Foundation (NCF) for the use of supercomputer facilities, with financial support from the Netherlands Organization for Scientific
Research (NWO) through VICI grant 639.043.409. The research was supported in part by the European Research Council under the European Union's Seventh 
Framework Programme (FP7/2007-2013) / ERC Grant agreements 278594-GasAroundGalaxies. This research was supported by ERC FP7 grant 279396 and ERC FP7 
grant 278594. H. Hildebrandt is supported by an Emmy Noether grant (No. Hi 1495/2-1) of the Deutsche Forschungsgemeinschaft.

MC thanks Alma and Sofia for their help in getting priorities straight.

\emph{Author Contributions:} All authors contributed to the development and writing of this paper. The authorship list is given in three groups: the lead authors (MVe, MC, HHo, JS), followed by two alphabetical groups. The first alphabetical group includes those who are key contributors to both the scientific analysis and the data products. The second group covers those who have either made a significant contribution to the data products, or to the scientific analysis.

\bibliographystyle{mnras}
\bibliography{ESD} 

\appendix

\section{Comparison of analytical and numerical \ESD profiles}

The measured galaxy-galaxy lensing signal is often interpreted in the context of a $\Lambda$CDM framework where the baryon content of a (lens) galaxy is embedded in a dark matter halo. The lensing effect on the light rays emitted by background (source) galaxies is therefore caused by the \emph{total} matter density contrast along the line of sight.
At the transverse separations of interest in this paper ($0.02<R<2 h^{-1}$ Mpc) most of this matter contrast is actually associated with the foreground (lens) galaxy. If one further limits the analysis to central galaxies, a simple yet effective model --often employed in the literature-- assumes that i) the contribution to the lensing signal from the stellar mass of a galaxy can be described as a point-mass contribution ($\Delta \Sigma_{\rm star} (R) \propto R^{-2}$), ii) the contribution from (both cold and hot) gas can be ignored ($\Delta \Sigma_{\rm gas} (R) \approx 0)$; and iii) the contribution from the dark matter halo, $\Delta \Sigma_{\rm halo}(R)$, can be described analytically assuming a spherical NFW matter density profile. The ESD profiles computed from the EAGLE simulation represent a benchmark against which the simple model outlined above can be tested. With this aim, we proceed with the following tests.

\subsection{ESD signal of the mean halo and stellar mass}
\begin{figure*} 
\begin{center} 
\includegraphics[width=2.0\columnwidth]{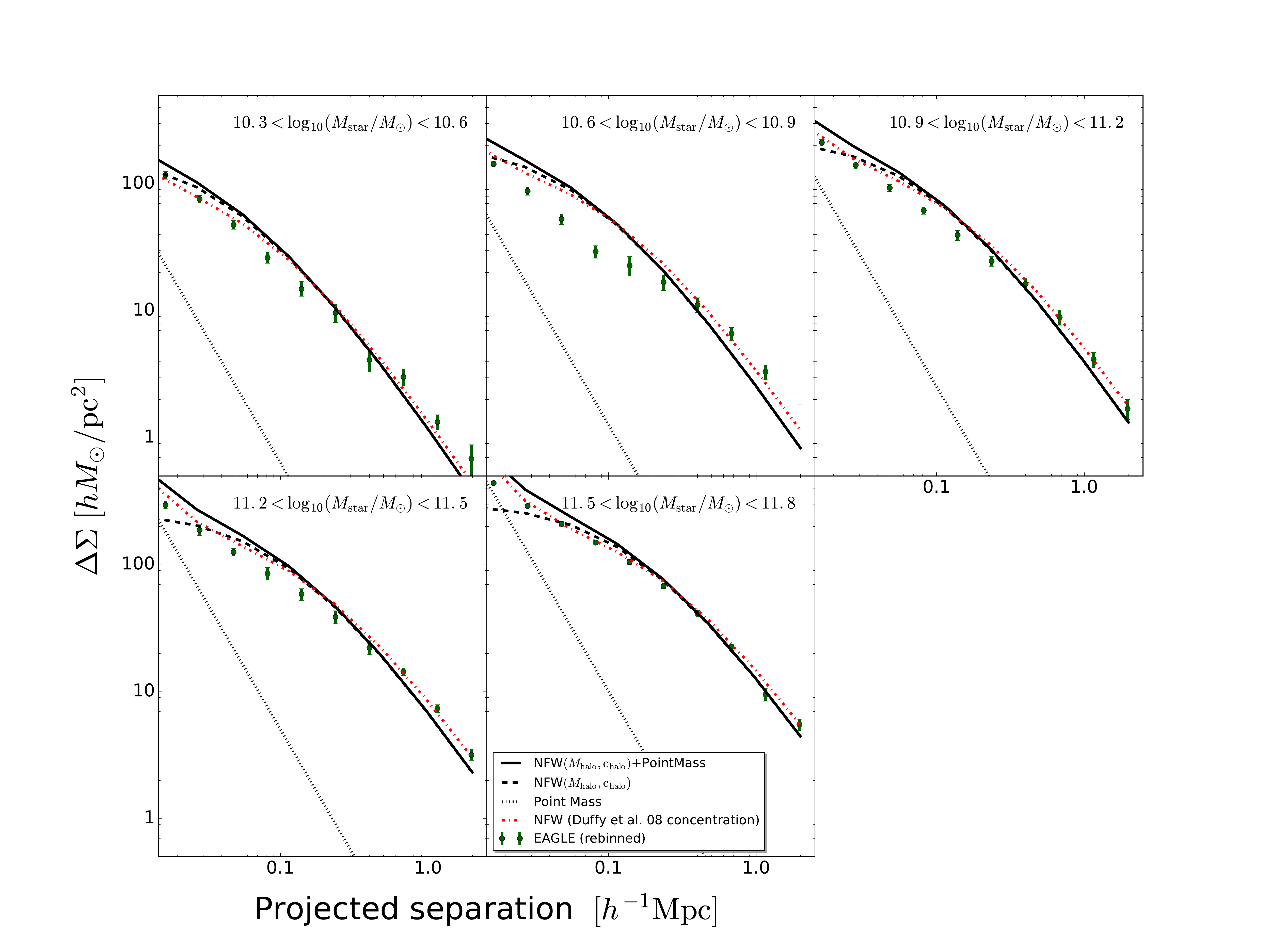} 
\end{center}
\caption{Excess surface density computed for central galaxies in EAGLE for different stellar mass bins (green circles with error bars). These profiles are the rebinned versions of those plotted with a red line in Figure~4. Analytical predictions of the excess surface are plotted with different line styles (see legend). For the point mass term, we use the mean stellar mass in each bin. For the NFW term, we use the mean halo mass for each bin as reported in column~2 of Table~1 in the main body of the paper and the corresponding halo concentration according to \citet{Schaller15}.} 
\label{figs:nfw_sim}
\end{figure*}

For each stellar mass bin for which \ESD is computed, we know which haloes contribute to the stack. We thus can compute the mean\footnote{We have repeated the exact test either using the median (stellar and halo) masses in the bins or using the entire distribution of (stellar and halo) masses and in both cases the results do not change significantly.} halo mass for each stellar mass bin (see column 2 in Table 1). We compute $\Delta \Sigma_{\rm halo}(R)$ corresponding to this mass adopting the concentration-halo mass relation  derived for (relaxed\footnote{We further comment on this in the next subsection.}) haloes of the EAGLE simulation  \cite[see][]{Schaller15}. We also compute $\Delta \Sigma_{\rm star} (R)$ using the mean stellar mass in each bin. The results are shown in Figure~\ref{figs:nfw_sim} where the ESD profiles of the simulations (green points with error bars) have been rebinned to 10 radial points, and the different contributions are indicated with different line styles and colours as indicated in the legend. 

The analytical description of the ESD profiles is in fair agreement for all stellar mass bins only on scales larger than $R\sim0.25 h^{-1}$Mpc. On smaller scales the analytical description systematically overestimates the results from the simulations. The agreement on relatively large scales suggests that the knowledge of the mean total mass of the halo is indeed sufficient to describe the lensing signal at those scales. On smaller scale, however, the ESD profile is clearly dependent on the actual matter density distribution. The fact that the simulations are systematically below the analytical predictions seems to indicate that the haloes that contribute to the signal are less centrally concentrated than what is assumed. \citet{Schaller15} show that the dark matter haloes in the EAGLE simulation have slightly different concentrations than those in the dark-matter only version of EAGLE. However, the difference is not sufficient to explain the feature under inspection here. It is worth noting that the concentration-mass relation provided by \citet{Schaller15} and adopted here for this test was derived using only relaxed haloes (for which a spherical NFW matter density profile is an adequate description). Not all of the haloes that enter the stack in each stellar mass bin are expected to be relaxed and this may be the cause of the difference between the analytical and numerical ESD profiles. \citet{Duffy08} reported indeed that, in the case of the OWLS simulations \citep{Schaye10}, a sample with only relaxed haloes yields on average higher concentrations than a sample where also unrelaxed haloes are included. We show that using the halo concentration-mass relation in \citet{Duffy08} for the full sample indeed leads to lower \ESD profiles on scales below $\sim 0.25 h^{-1}$Mpc. This in turn yields a better agreement with the profiles predicted from the EAGLE simulation (dot-dashed red line in Figure~A1). Despite the improvement, significant differences are still noticeable for the four lowest mass bins and on scales $0.03<R/(h^{-1}$ Mpc$)<0.2$. We defer a more quantitative analysis of this feature to further publications as it is beyond the scope of this paper.  

\subsection{Fitting the numerical \ESD profiles}

The test described in \S~A1 shows that a simple analytical model cannot reproduce the entire scale dependence of the ESD profiles obtained from the EAGLE simulation. The question then arises whether this severely hampers the possibility to retrieve halo properties such as their masses and concentrations when such simple models are employed to fit the ESD profiles. To answer this question we define a model in which 
\begin{equation}
\Delta \Sigma (R) = \Delta \Sigma_{\rm star} (R|\langle M_{\rm star}\rangle) + \Delta \Sigma_{\rm halo}(R|M_{\rm halo}, {\rm c}_{\rm halo}) \, .    
\end{equation}

\begin{figure*} 
\begin{center} 
\includegraphics[width=2.0\columnwidth]{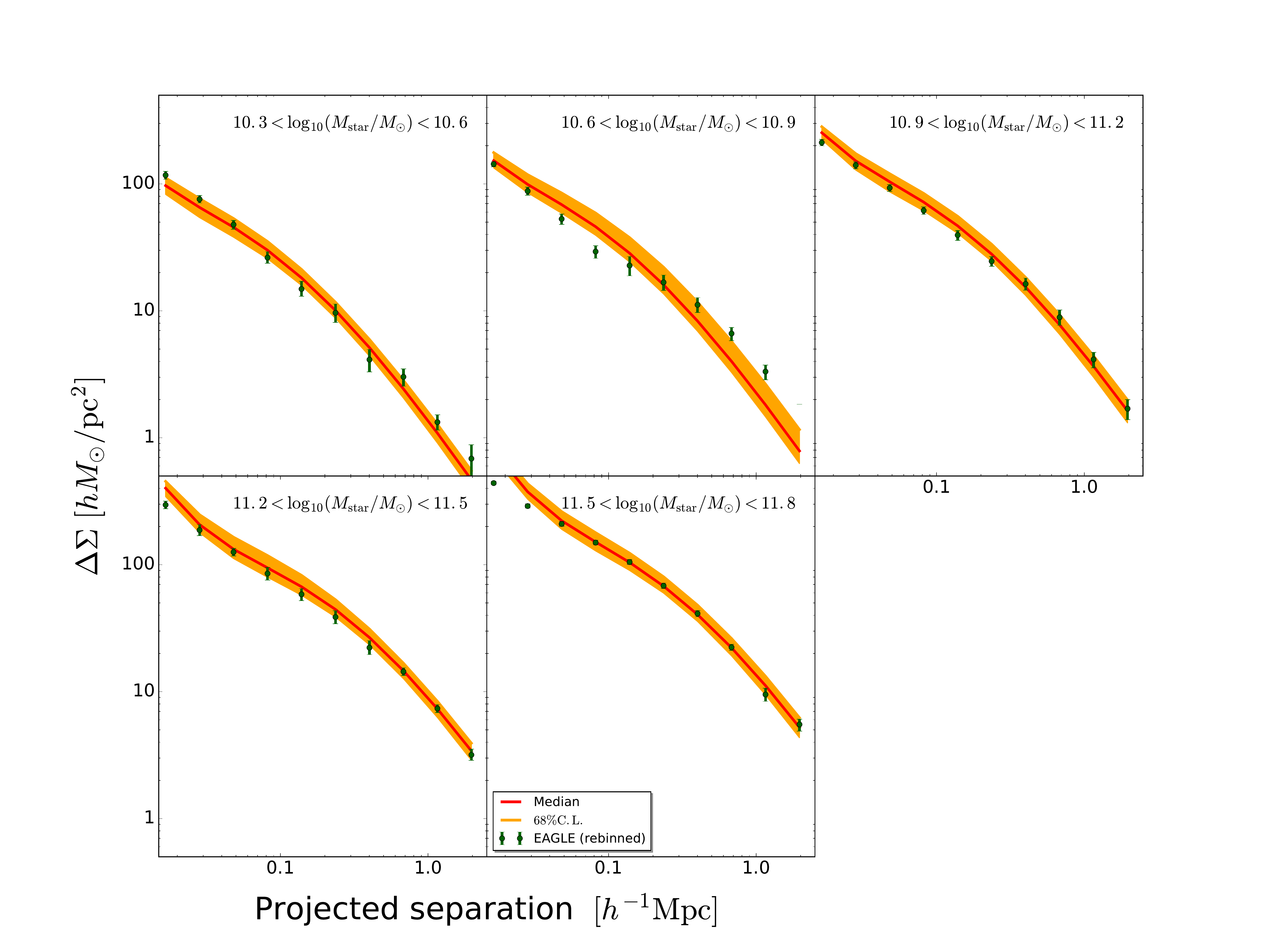} 
\end{center}
\caption{Excess surface density computed for central galaxies in EAGLE for different stellar mass bins (green circles with error bars, these profiles are the rebinned versions of those plotted with a red line in Figure~4.). Red curves (and orange shades) represent the median (and the 68\% credibility intervals) derived from the MCMC employed to fit the data. A fair description of the data can be obtained for each bin except for that corresponding to galaxies in the stellar mass range $10.6<\log(M_{\rm star}/M_{\odot})<10.9$ (see text for a discussion about this feature).} 
\label{figs:fit_sim}
\end{figure*}

Here, $\langle M_{\rm star}\rangle$ is a free parameter that indicates the mean stellar mass in each stellar mass bin\footnote{We adopt the same stellar mass bins as in the main body of the paper.}, $M_{\rm halo}$, and  ${\rm c}_{\rm halo}$ are two free parameters that completely specify the analytical ESD profile of a halo with a NFW matter density profile. We treat the ESD profiles from the EAGLE simulation as the data to be fit by this model. We fit $\langle M_{\rm star}\rangle$, $M_{\rm halo}$, and  ${\rm c}_{\rm halo}$ independently for each stellar mass bin. No priors are imposed on $M_{\rm halo}$, and  ${\rm c}_{\rm halo}$, whereas we impose that $\langle M_{\rm star}\rangle$ is within the stellar mass bin limits. The fit is performed using a Markov Chain Monte Carlo (MCMC) technique. Specifically, we employ the publicly available emcee code \citep{emcee} and we check for convergence by ensuring that the chain is much longer than the auto-correlation length of each parameter. We find that the best-fit model yields a $\chi^2_{\rm red} = 48.9/(50-10) = 1.22$, i.e. the simple model can adequately describe the data, although more flexible models might yield even better agreement. Figure A2 shows the ESD profiles from the simulation (green points with error bars) and the median and the 68\% credibility level (red curves and orange shaded regions) derived from the MCMC run. 

\begin{figure} 
\begin{center} 
\includegraphics[width=\columnwidth]{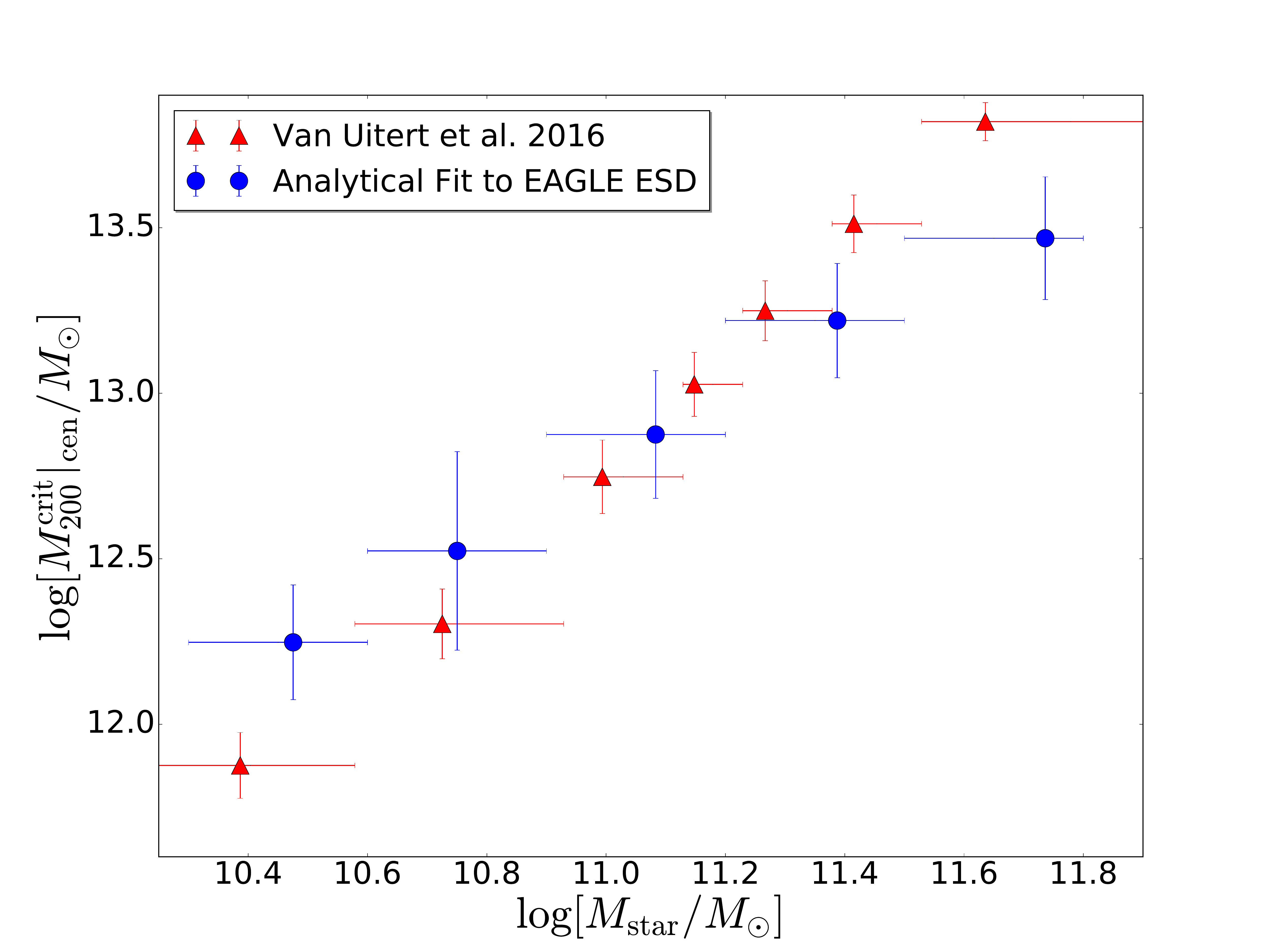} \\
\includegraphics[width=\columnwidth]{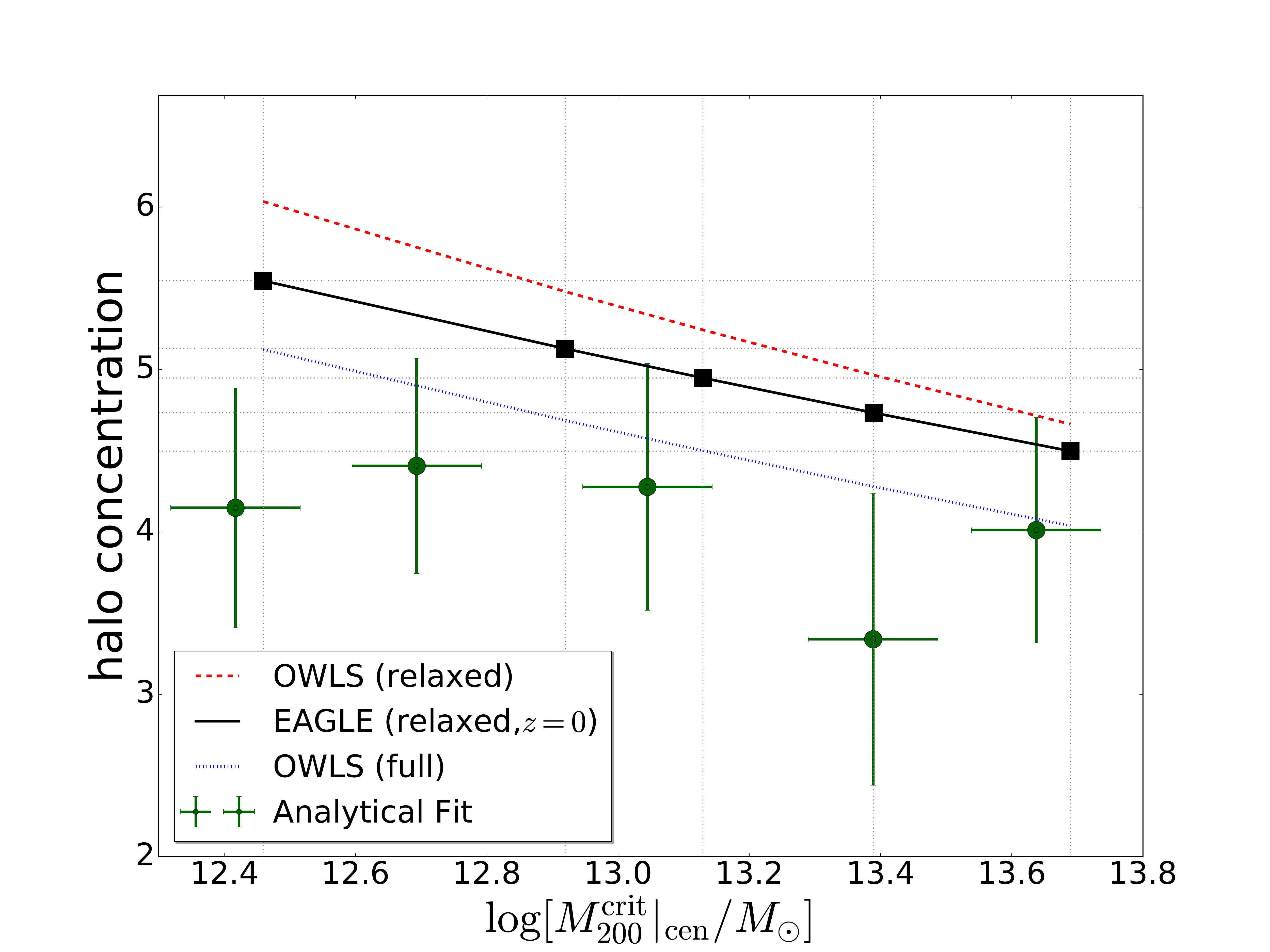}
\end{center}
\caption{
{\it Top Panel.} 
Average halo mass (for haloes with $N_{\rm FoF}^{\rm GAMA} \ge 5$) of central galaxies binned in stellar mass. Comparison between the result of the MCMC run on the EAGLE \ESD profile (see detaill in \S~A2) and the result from \citet{vanUitert16} obtained simultaneously fitting galaxy-galaxy lensing and the GAMA stellar mass function for the KiDSxGAMA galaxy sample. Both blue and red horizontal bars indicate the width of the bin, whereas vertical bars indicate the 68\% credibility interval for the inferred halo mass.
{\it Bottom Panel.} 
Halo concentration-mass relation. Green circles with error bars refer to the median and the 68\% credibility interval obtained from the MCMC used to fit the EAGLE \ESD profiles. The black line represents the relation for relaxed haloes in the EAGLE simulation \citep[see][]{Schaller15}, where the black points indicate the mean halo masses of the five stellar mass bins used in the analysis. For reference, the concentration mass relations from \citet{Duffy08} are also reported with red dashed and blue dotted lines, indicating the relation derived for relaxed-only and all haloes, respectively (see discussion in \S A2).
}
\label{figs:c_m_fit}
\end{figure}

The top panel of Figure~\ref{figs:c_m_fit} shows the constraints on the mean stellar and halo masses obtained by fitting the \ESD of the EAGLE  simulation with the simple analytical model described above. For comparison, we report the results by \cite{vanUitert16} obtained simultaneously fitting the KiDSxGAMA galaxy-galaxy lensing profile and the GAMA stellar mass function. We note here that  \cite{vanUitert16} employed a sophisticated halo model rather than a simple three-parameter (per bin) model like the one adopted here. In the range, $10.6<\log[M_{\rm star}/M_{\odot}]<11.5$, we find a very satisfactory agreement between the two stellar-halo mass relations. We note, however, that the inferred halo masses for the lowest and the highest stellar mass bins differ at 2-sigma level. In this paper, we refrain from investigating the source of this disagreement in any detail but we defer to Figure~3 of \cite{vanUitert16} for a discussion of the differences in the constraints on the halo masses when one fits galaxy-galaxy lensing alone or jointly with the stellar mass function.

The bottom panel of Figure~\ref{figs:c_m_fit} shows the constraints on the halo concentration and mass obtained by fitting the \ESD of the EAGLE  simulation with the simple analytical model described above. As expected from the test in \S~A1 the posterior distributions of the parameters $M_{\rm halo}$ and  ${\rm c}_{\rm halo}$ indicate that concentrations are systematically underestimated with respect to the fiducial concentration-halo mass relation (black points in Figure~\ref{figs:c_m_fit}). The result we find confirms the notion that the halo concentrations found via a fitting of the ESD profiles have to be interpreted as \emph{effective} concentrations and are most likely to be lower than those based on fits of relaxed haloes in numerical simulations. This has already been noted in several observational works, e.g. in the context of fitting ESD profiles around GAMA galaxies using KiDS galaxy images \citep[see e.g.][]{Viola15,vanUitert16}.
A closer inspection of Figure~A3 also shows that the retrieved mean halo masses (circles with horizontal error bars) are unbiased with respect to the actual mean halo mass in each stellar mass bin in all cases except for the bin $10.6<\log_{10}[M_{\rm star}/M_{\odot}]<10.9$. This is perhaps not surprising given that this is exactly the bin for which the ESD profile from the EAGLE simulation differs the most from an analytical \ESD profile that assumes a NFW matter density profile. 

Finally, we note that, at the smallest scales probed here ($R<0.03 h^{-1}$Mpc), the \ESD profile is sensitive to the point mass assumption employed to describe the contribution from the stellar content of the galaxy. The simulation disfavours a steep profile $\Delta \Sigma(R) \propto R^{-2}$ and a better fit at those scales would require a more detailed description of the stellar mass distribution in galaxies \citep[see e.g.][for a similar discussion in the context of the galaxy-galaxy lensing quality in forthcoming surveys]{Kobayashi15}.

\bsp	
\label{lastpage}
\end{document}